\documentclass[12pt]{article}
\usepackage{amsmath,amssymb,bbold,epsfig}
\usepackage{amsfonts}
\usepackage{slashed}
\usepackage{dsfont}
\usepackage{cite}
\usepackage{graphicx,color}
\usepackage{bbm}
\usepackage[usenames,dvipsnames]{xcolor}
\definecolor{lightgray}{gray}{0.90}
\usepackage[colorlinks=true,linkcolor=red,citecolor=green,urlcolor=cyan,
bookmarks=true,bookmarksopen=true,pdfpagemode=None,pdfstartview=FitH]{hyperref}
\newcommand{\appsection}{\addtocounter{section}{1}\setcounter{equation}{0}
                         \renewcommand{\thesection}{\Alph{section}}
}
\renewcommand{\theequation}{\arabic{equation}}
\newcommand{\be}{\begin{equation}}
\newcommand{\ee}{\end{equation}}
\newcommand{\bea}{\begin{eqnarray}}
\newcommand{\eea}{\end{eqnarray}}
\begin{document}

\title{\vspace{-2cm}
\vglue 1.3cm
\Large \bf
Another look at synchronized neutrino oscillations}
\author{
{Evgeny~Akhmedov$^1$\footnote{Also at the National Research Centre 
Kurchatov Institute, Moscow, Russia}~\thanks{Email: \tt 
akhmedov@mpi-hd.mpg.de}
\;and \,Alessandro Mirizzi$^{2,3}$\thanks{Email: \tt  
alessandro.mirizzi@ba.infn.it}
\vspace*{3.5mm}
} \\
{\normalsize\em 
$^1$Max-Planck-Institut f\"ur Kernphysik, Saupfercheckweg 1, \vspace*{-1mm}}\\
{\normalsize\em
69117 Heidelberg, Germany
} 
\\
{\normalsize\em 
$^2$Dipartimento Interateneo di Fisica ``Michelangelo Merlin'',
\vspace*{-1mm}
} \\
{\normalsize\em 
Via Amendola 173, 70126 Bari, Italy 
\vspace*{-0.60cm}
}
\\
{\normalsize\em 
} \\
{\normalsize\em 
$^3$Istituto Nazionale di Fisica Nucleare - Sezione di Bari,
\vspace*{-1mm}} \\
{\normalsize\em 
Via Amendola 173, 70126 Bari, Italy \vspace*{0.15cm}}
}
\date{}  

\maketitle
\thispagestyle{empty}
\vspace{-0.8cm}
\begin{abstract}
In dense neutrino backgrounds present in supernovae and in the early 
Universe neutrino oscillations may exhibit complex collective phenomena, 
such as synchronized oscillations, bipolar oscillations and spectral 
splits and swaps. We consider in detail possible decoherence effects on 
the simplest of these phenomena -- synchronized neutrino oscillations 
that can occur in a uniform and isotropic neutrino gas. We develop an 
exact formalism of spectral moments of the flavour spin vectors 
describing such a system and then apply it to find 
analytical 
approaches that allow one to study decoherence effects on its late-time 
evolution. This turns out to be possible in part due to the existence of 
the (previously unknown) exact conservation law satisfied by the 
quantities describing the considered neutrino system. Interpretation of 
the decoherence effects in terms of neutrino wave packet separation is 
also given, both in the adiabatic and non-adiabatic regimes of neutrino 
flavour evolution.  
\end{abstract} 
\vspace{0.2cm} 
\centerline{Pacs numbers: 14.60Pq, 97.60Bw} 
\vspace{0.3cm} 

\newpage


\section{\label{sec:intro}Introduction}

It is well known that neutrino oscillations in dense neutrino backgrounds 
existing at certain stages of supernova explosion and in the early 
Universe may differ drastically  from the oscillations in ordinary matter or 
in vacuum. In particular, synchronized oscillations \cite{synch1,synch2,Pantaleone:1992eq,synch3,
synch4,synch5,synch6,RaffTamb}, bipolar oscillations \cite{bipolar,synch4,duan1,Duan:2006an,
hann1,Fogli:2007bk,duan2}, spectral splits and swaps \cite{splits1,splits2,splits3,splits4} 
and multiple spectral splits \cite{multisplits} are possible. These phenomena 
have attracted a great deal of attention recently, see 
Refs.~\cite{review1,review2} for reviews and extensive lists of literature.  

The simplest system that exhibits collective neutrino oscillations is a dense 
uniform and isotropic gas consisting of only neutrinos (or only antineutrinos). 
In such a system, for sufficiently large neutrino density, neutrinos of 
different energies oscillate with the same frequency (i.e.\ undergo 
synchronized oscillations), and therefore even for wide neutrino spectra 
the oscillations do not average out with time. This is in sharp contrast with 
what is expected in the case of neutrino oscillations in vacuum, in usual 
matter or in low-density neutrino backgrounds. In particular, in vacuum 
neutrinos of different energies oscillate with different frequencies and over 
the time develop large phase differences, leading to decoherence and averaging 
out of the oscillations. On the other hand, synchronized neutrino oscillations 
in a dense neutrino gas mean that no decoherence occurs (or at least that some 
degree of coherence is maintained) in such a system, since complete 
decoherence would destroy the synchronization.

In this paper we explore late-time decoherence effects on 
collective neutrino oscillations. To this end, we concentrate on the simplest 
possible system where collective oscillations can take place --  
a uniform and isotropic neutrino gas. 
Decoherence of neutrino oscillations 
can be described either in momentum space or in coordinate 
space. In the momentum space it comes from the dephasing of different neutrino 
modes at late times and is related to the integration over the neutrino
spectrum. In the coordinate space decoherence is related to the spatial
separation of the wave packets of different neutrino propagation 
eigenstates after they have traveled long enough distance. The momentum-space 
and coordinate-space descriptions are equivalent (see, e.g., \cite{paradoxes}).

Since in supernovae and in the early Universe neutrinos are produced at 
very high densities, their production processes are well localized in 
space and time and therefore their wave packets are very short in 
coordinate space \cite{Kersten,KerstSmir,AKL4}. As a result, one could 
expect decoherence by wave packet separation to occur rather quickly and 
to affect significantly collective neutrino oscillations. In particular, 
this would destroy synchronized neutrino oscillations at sufficiently 
late times. Numerical calculations show, however, no trace of such 
decoherence when the density of the neutrino gas is high enough.  One of 
the main goals of the present study was therefore to understand why no 
decoherence (and therefore no de-synchronization) occurs in high-density 
neutrino gases, and how in general coherence and decoherence are related 
to the synchronization of neutrino oscillations or lack thereof.
Our study is in a sense complementary to that in \cite{synch3} 
where the possibility for a neutrino system to develop a spontaneous
synchronization starting with a completely incoherent initial state 
was considered.

The paper is organized as follows. In Section~\ref{sec:flspin} we review 
the standard flavour spin formalism which is especially well suited for 
describing neutrino oscillations and flavour conversions in dense 
neutrino backgrounds. We also discuss the conservation law for a 
quantity ${\cal E}$ which can be interpreted as the total energy of 
self-interacting magnetic moments in an external magnetic field. This 
section mainly serves to introduce our framework and notation. 
Sections~\ref{sec:specmoments} - \ref{sec:interp} contain our new 
results. In Section~\ref{sec:specmoments} we develop a formalism of 
spectral moments $\vec{K}_n$ describing a homogeneous and isotropic gas 
of neutrinos or neutrinos and antineutrinos. We derive equations of 
motion for these quantities and relations between the time derivatives 
of $\vec{K}_n$ and $\vec{K}_{n+1}$. We also establish a new conservation 
law for this neutrino system, not previously known in the literature. In 
Section~\ref{sec:latetime} we develop two approximate analytical 
approaches for describing decoherence effects on synchronized neutrino 
oscillations. They are based on the formalism of Section 3 augmented by 
certain assumptions about the late-time behaviour of the neutrino 
flavour spin vector in the system under consideration. In this section 
we also compare our approach and its results with those in 
Ref.~\cite{RaffTamb}, where decoherence effects on synchronized neutrino 
oscillations have also been studied. In Section~\ref{sec:interp} we give 
a qualitative interpretation of coherence and partial or full 
decoherence in terms of wave packet separation based on the 
consideration in the neutrino propagation eigenstate basis. The roles of 
adiabaticity and adiabaticity violation for possible decoherence effects 
is considered. In Section~\ref{sec:summary} we summarize and discuss our 
results. Technical details of some derivations related to our analysis 
in Section \ref{sec:latetime} are given in the Appendix.

\section{\label{sec:flspin}The flavour spin formalism}

Flavor mixing and evolution in a neutrino gas can be described by 
time-dependent density matrices ${\varrho}_{\bf p}$, which for each 
momentum mode ${\bf p}$ are matrices in flavour space 
\cite{Dolgov,BarbDolg,RaffSigl}. Their diagonal elements are actually 
occupation numbers for neutrinos of given flavour, while the 
off-diagonal elements contain information about coherence properties of 
the system. The evolution of these matrices is governed by the Liouville 
equation~\cite{RaffSigl,McKellar:1992ja,Cardall:2007zw,Vlasenko:2013fja}.

We will consider a homogeneous and isotropic neutrino gas evolving in time. 
It has been recently realized that even very small initial deviations from 
space-time symmetries of a system of self-interacting neutrinos may be 
strongly enhanced in the course of its evolution \cite{Raffelt:2013rqa,
Duan:2014gfa,Mirizzi:2015fva}. Such effects 
could profoundly influence the flavor evolution of the system and are 
currently under active investigation. Here we ignore such complications 
and assume that the uniformity and isotropy of the neutrino gas are exact 
and are preserved during its evolution.

For simplicity, we confine ourselves to 2-flavour neutrino oscillations 
$\nu_e\leftrightarrow \nu_x$, where $\nu_x = \nu_\mu,\,\nu_\tau$ or a 
superposition thereof. For an isotropic 
system one can use the absolute value of the neutrino momentum 
$p\equiv |{\bf p}|$ rather than the momentum itself to label the neutrino 
kinematic characteristics. 
However, it is more convenient to use instead the vacuum oscillation frequency
\be
\omega=\frac{\Delta m^2}{2p}\,,
\label{eq:omega}
\ee
with $\Delta m^2$ being the mass squared difference of neutrino mass 
eigenstates. In the 2-flavour case one can 
decompose the density matrices and the Hamiltonian in terms of the Pauli 
matrices $\sigma_i$. The flavour evolution of each $\omega$-mode can 
then be described by the equation of motion (EoM) of the corresponding 
flavour spin vector \cite{RaffSigl}:
\be
\dot{\vec{P}}_\omega=\vec{H}_\omega\times\vec{P}_\omega\,.
\label{eq:EoM1}
\ee
Here $\vec{H}_\omega$ is the Hamiltonian (or ``effective magnetic field'')  
vector:
\be
\vec{H}_\omega=\omega\vec{B}+\lambda\vec{L}+\mu \vec{P}\,
\label{eq:vecH1}
\ee 
with
\be
\vec{B}=(s_{20},\,0,\,-c_{20})\,,\qquad \vec{L}=\vec{n}_z\equiv(0,\, 0,\, 1),
\qquad\lambda=\sqrt{2}G_F n_e\,,\qquad
\mu=\sqrt{2}G_F n_\nu\,.
\label{eq:omegaBmu}
\ee
Here $s_{20}\equiv\sin 2\theta_0$, $c_{20}\equiv\cos 2\theta_0$ with 
$\theta_0$ being the leptonic mixing angle in vacuum, $G_F$ is the Fermi 
constant, and $n_e$ and $n_\nu$ are the net electron and  
neutrino number densities, respectively (i.e.\ the differences of number 
densities of the corresponding particles and antiparticles). The quantity 
$\vec{P}$ is the global flavour spin vector of the neutrino system: 
\be
\vec{P}=\int_{-\infty}^\infty d\omega \vec{P}_\omega\,. 
\label{eq:D}
\ee 
We use the convention according to which positive values of $\omega$ correspond 
to neutrinos and negative values to antineutrinos \cite{RaffSigl,review1}. 
The terms in eq.~(\ref{eq:vecH1}) proportional to $\omega$, $\lambda$ 
and $\mu$ correspond, respectively, to the vacuum contribution to the 
neutrino Hamiltonian and to the contributions coming from coherent 
forward scattering of a test neutrino on the particles of ordinary 
matter and on the neutrino background. If the density of ordinary matter 
is constant or nearly constant in the region where collective neutrino 
oscillations are expected to take place, effects of ordinary matter can 
be removed by going into a frame rotating around $\vec{L}$ and replacing 
$\theta_0$ by an effective mixing angle \cite{duan1,hann1}. In what follows 
we will be assuming that this has already been done (or that the effects of 
ordinary matter are negligible), and we will keep the notation $\theta_0$ 
for the mixing angle defining the vector $\vec{B}$. The vector 
$\vec{H}_\omega$ can then be written as
\be
\vec{H}_\omega=\omega\vec{B}+\mu \vec{P}\,.
\label{eq:vecH2}
\ee 

Equations of motion (EoMs) (\ref{eq:EoM1}) describe the precession of the 
flavour spin vectors $\vec{P}_\omega$ of the individual neutrino modes around 
their corresponding ``magnetic field'' vectors $\vec{H}_\omega$. Obviously, 
they conserve the lengths of $\vec{P}_\omega$:
\be
|\vec{P}_\omega|\equiv P_0 g_\omega=const.
\label{eq:g}
\ee 
The function $g_\omega$ is just the spectrum of neutrinos in the variable 
$\omega$, which we will assume to be normalized according to 
\be
\int g_\omega d\omega = 1\,
\label{eq:norm1}
\ee 
and to have the effective width $\sigma_\omega$. For example, for the Gaussian 
spectrum 
\be
g_\omega=\frac{1}{\sqrt{2\pi}\sigma_\omega}
e^{-\frac{(\omega-\omega_0)^2}{2\sigma_\omega^2}}\,.
\label{eq:Gauss}
\ee 
In our study we will be assuming that $g_\omega$ corresponds to the 
$\omega$-spectrum of the wave packets of individual neutrinos, i.e.\ 
we deal with an ensemble of neutrinos described by identical wave packets  
with the same mean energy. 
Generalizations to more general neutrino spectra is straightforward.    
Note that a system of wave packets with the energy distribution 
function $g_\omega$ is equivalent to a system of neutrinos with well defined 
energies and spectrum $g_\omega$~\cite{Kiers}, 
to which our treatment will therefore also apply.

The flavour content of a given $\omega$-mode is defined by the projection of 
$\vec{P}_\omega$ on the $z$-axis in the flavour space; 
for $\nu_e$ and  $\bar{\nu}_x$ this projection is positive, whereas for 
$\bar{\nu}_e$, $\nu_x$ it is negative. For systems consisting initially of 
arbitrary numbers of the flavour eigenstate neutrinos $\nu_e$ and $\nu_x$ and 
of their antineutrinos the initial conditions for the individual $\omega$-modes 
are therefore 
\be
\vec{P}_\omega(0)=P_0 g_\omega\vec{n}_z\,.
\label{eq:init1}
\ee 
The initial condition for the global flavour spin vector is then 
\be
\vec{P}(0)=P_0 \vec{n}_z\,.
\label{eq:init2}
\ee 
The parameter $P_0$ (i.e.\ the initial length of the vector 
$\vec{P}$) could in principle be set equal to one through a proper redefinition 
of the neutrino self-interaction strength $\mu$. However, in certain 
situations it may be convenient to define $\mu$ as being proportional to the 
number density of just one neutrino species (e.g., 
$\bar{\nu}_e$ \cite{review1}). We therefore keep $P_0$ as a free parameter. 

Integrating eq.~(\ref{eq:EoM1}) over $\omega$, we obtain the EoM for $\vec{P}$:
\be
\dot{\vec{P}}=\vec{B}\times\vec{S}\,,\qquad \text{where}\qquad 
\vec{S}\equiv \int d\omega \omega \vec{P}_\omega\,.
\label{eq:EoM2}
\ee
{}From the conservation of $|\vec{P}_\omega|$ and the initial condition 
(\ref{eq:init2}) it follows that for $t>0$ the length of the vector $\vec{P}$ 
satisfies $P(t)\equiv|\vec{P}(t)|\le P_0$. At the same time, 
eq.~(\ref{eq:EoM2}) implies 
\be
\vec{P}\!\cdot\!\vec{B}=const.=\vec{P}(0)\!\cdot\!\vec{B}=-c_{20}P_0\,.
\label{eq:PB}
\ee
Therefore,  
\be
c_{20}P_0\le P(t)\le P_0\,.
\label{eq:ineq1}
\ee

{}From the EoMs (\ref{eq:EoM1}) and (\ref{eq:EoM2}) it follows that 
the following quantity is an integral of motion \cite{duan1,hann1}:
\be
{\cal E}\equiv \vec{B}\!\cdot\!\vec{S}+\frac{\mu P^2}{2}=const.
\label{eq:Etot1}
\ee
It can be interpreted as the total energy  
of a system of `spins' $\vec{P}_\omega$ with magnetic moments characterized by 
`gyromagnetic ratios' $\omega_i$. The quantity $\vec{S}$ is then the total 
magnetic moment of the system. In this interpretation, the first term in 
(\ref{eq:Etot1}) describes the interaction of the spins with the `external 
magnetic field' $\vec{B}$, whereas the second term describes the spin-spin 
interaction \cite{duan1,hann1,synch4}.  
The system of self-interacting neutrinos is thus mathematically equivalent 
to the system of classical magnetic dipoles with the Hamiltonian 
given by eq.~(\ref{eq:Etot1}). Many properties of the former can therefore be 
understood from the properties and symmetries of the latter \cite{Nmode}. 
The initial conditions (\ref{eq:init1}) imply that 
\be
\vec{S}(0)=\omega_0 P_0\vec{n}_z\,, \qquad\text{where}\qquad 
\omega_0=\langle\omega\rangle\equiv\int g_\omega \omega d\omega\,. 
\label{eq:S0}
\ee 
Here $\omega_0$ is the mean neutrino `energy' corresponding to the spectrum 
$g_\omega$. For symmetric spectra (such as e.g. the Gaussian spectrum 
(\ref{eq:Gauss})) $\omega_0$ coincides with the frequency at which $g_\omega$ 
reaches its peak value. Substituting eqs.~(\ref{eq:init2}) and (\ref{eq:S0}) 
into (\ref{eq:Etot1}), we find 
\be
\frac{\mu P^2}{2}+\vec{B}\!\cdot\!\vec{S}=\frac{\mu P_0^2}{2}-
c_{20}\omega_0P_0\,.
\label{eq:Etot2}
\ee

\section{\label{sec:specmoments}The formalism of spectral moments} 

Let us introduce the spectral moments $\vec{K}_n(t)$ of the flavour spin 
vector: 
\be
\vec{K}_n(t) = \int d\omega\,\omega^n \vec{P}_{\omega}(t)\,,\qquad n\ge 0\,.
\label{eq:Kn}
\ee
The integral on the right hand side of (\ref{eq:Kn}) is well defined as far as 
the neutrino spectrum $g_\omega$ goes to zero fast enough for 
$|\omega|\to\infty$ (this is the case e.g.\ for the Gaussian spectrum and for 
any spectrum which vanishes outside a finite interval of $\omega$, such as the 
box-type spectrum). Note that  the quantities $\vec{P}$ and $\vec{S}$ 
discussed in Section~\ref{sec:flspin} are just particular cases of the 
spectral moments:
\be
\vec{P}(t)=\vec{K}_0(t)\,,\qquad 
\vec{S}(t)=\vec{K}_1(t)\,.
\label{eq:K0K1}
\ee
Rewriting eq.~(\ref{eq:EoM1}) as $\dot{\vec{P}}_\omega=(\omega\vec{B}+\mu 
\vec{P})\times\vec{P}_\omega$, multiplying by $\omega^n$ and integrating 
over $\omega$, we obtain 
\be
\dot{\vec{K}}_n = \vec{B}\times\vec{K}_{n+1}+\mu\vec{P}\times 
\vec{K}_n
\,.
\label{eq:Kndiff1}
\ee
{}From this equation it is straightforward to find a scalar relation between  
the derivatives of $\vec{K}_n$ and $\vec{K}_{n+1}$:%
\footnote{Alternatively, one can derive eq.~(\ref{eq:Kndiff2}) by noting that 
eq.~(\ref{eq:EoM1}) implies $\vec{H}_\omega\!\cdot\!\dot{\vec{P}}_\omega=0$. 
Multiplying this relation by $\omega^n$ and integrating over $\omega$ 
immediately yields eq.~(\ref{eq:Kndiff2}).}
\be
\vec{B}\!\cdot\!\dot{\vec{K}}_{n+1}+\mu\vec{P}\!\cdot\!\dot{\vec{K}}_n=0
\,.
\label{eq:Kndiff2}
\ee
For $n=0$ this gives $\vec{B}\!\cdot\!\dot{\vec{S}}+\mu\vec{P}\!\cdot\!
\dot{\vec{P}}=0$, which can be immediately integrated. The result is the 
already known conservation law for 
${\cal E}$, eq.~(\ref{eq:Etot1}). For $n=1$ we obtain 
\be
\vec{B}\!\cdot\!\dot{\vec{K}}_{2}+\mu\vec{P}\!\cdot\!\dot{\vec{S}}=0
\,.
\label{eq:Kndiff3}
\ee
Let us show that this relation can also be integrated. Indeed, from 
eq.~(\ref{eq:EoM2}) it follows that $\vec{S}\!\cdot\!\dot{\vec{P}}=0$. 
Therefore, $\vec{P}\!\cdot\!\dot{\vec{S}}=(d/dt)(\vec{P}\!\cdot\!\vec{S})$. 
This allows us to integrate eq.~(\ref{eq:Kndiff3}), which yields another 
conservation law satisfied by a system of self-interacting neutrinos:%
\footnote{
Alternatively, one can derive (\ref{eq:cons2}) by multiplying 
eq.~(4.10) of Ref.~\cite{baha1} by $\omega^2$ and integrating it \mbox{over 
$\omega$.} We thank Baha Balantekin for this comment.} 
\be
\tilde{\cal E}\equiv 
\vec{B}\!\cdot\!\vec{K}_{2}+\mu\vec{P}\!\cdot\!\vec{S}=const.
\label{eq:cons2}
\ee
While eq.~(\ref{eq:Etot1}) is well known, the conservation law (\ref{eq:cons2}) 
is new. The constant on its right hand side can be readily found if one 
observes that the initial conditions 
(\ref{eq:init1}) imply  
\be
\vec{K}_n(0)=P_0\langle \omega^n\rangle \vec{n}_z\,, \quad \mbox{where}
\quad \langle \omega^n\rangle \equiv \int\,d\omega\,\omega^n g_\omega\,. 
\label{eq:init3}
\ee
One can then rewrite eq.~(\ref{eq:cons2}) as 
\be
\vec{B}\!\cdot\!\vec{K}_{2}+\mu\vec{P}\!\cdot\!\vec{S}=P_0[\mu P_0\omega_0-
c_{20}(\omega_0^2+\sigma_\omega^2)]
\,.
\label{eq:cons2a}
\ee
Here we have used the relation $\langle\omega^2\rangle=\omega_0^2+
\sigma_\omega^2$, which is just the definition of the variance 
$\sigma_\omega^2$ of the $\omega$-spectrum. 

\section{\label{sec:latetime}Late-time regime of collective neutrino 
oscillations}
\subsection{\label{sec:previous}Previous studies}

Oscillations in a dense uniform and isotropic neutrino gas have been 
extensively studied 
in the literature (see, e.g., \cite{synch1,synch2,synch3,synch4,synch5,
synch6,RaffTamb}). 
However, to the best of 
 our knowledge, the only paper that dealt with the 
late-time decoherence effects on synchronized neutrino oscillations was 
Ref.~\cite{RaffTamb}).%
\footnote{Decoherence effects on flavour transformations of supernova 
neutrinos was also studied in Ref.~\cite{KerstSmir}. However, 
the influence of the coherence loss on collective neutrino oscillations 
has not be considered there.}
Here we briefly review its main results. 

The authors studied decoherence effects 
numerically and (under certain assumptions) analytically. As particular 
examples, two different neutrino spectra were considered: the Gaussian 
spectrum of unit variance ($\sigma_\omega=1$) and the box-type spectrum of 
overall width 2 (in units of some fiducial frequency). As an order parameter 
characterizing decoherence, the deviation of the transverse (with respect to 
the vector $\vec{B}$) component of the global flavour spin vector $\vec{P}$ 
from its initial value was taken. This choice, first suggested in 
\cite{synch3}, is well justified: Indeed, decoherence leads to a shrinkage 
of $\vec{P}$.%
\footnote{Note that, although the lengths of the vectors $\vec{P}_\omega$ of 
the individual modes are conserved by their EoMs, the length of $\vec{P}$ is 
in general not conserved.}
Since the longitudinal (with respect to $\vec{B}$) component of 
$\vec{P}$, $\vec{P}_\parallel\equiv (\vec{P}\!\cdot\!\vec{B})\vec{B}$,  
is conserved, only the transverse component $\vec{P}_\perp\equiv 
\vec{P}-(\vec{P}\!\cdot\!\vec{B})\vec{B}$ can shrink as a result of 
decoherence. A convenient choice for the order parameter is therefore 
$R_A\equiv P_\perp/P_\perp(0)=P_\perp/(s_{20}P_0)$; 
$R_A=1$ would then correspond to perfect coherence, whereas $R_A=0$ would 
imply complete decoherence \cite{synch3}. 

The authors of \cite{RaffTamb} found three regimes of neutrino oscillations 
in the system, depending on the value of neutrino self-interaction parameter 
$\mu$. 
\begin{itemize}
\item[(1)]
Large $\mu$ regime ~--~ ``perfect synchronization''.
Despite differences in $\omega$, all $\vec{P}_\omega$ evolve in a synchronized 
way, 
starting 
practically immediately from time $t=0$. The flavour spin vector $\vec{P}$ 
exhibits a simple precession around $\vec{B}$ with the frequency $\omega_0$ 
equal to the mean frequency of the neutrino spectrum:
\be
\dot{\vec{P}}=\omega_0 \vec{B}\times \vec{P}\,.
\label{eq:PPP}
\ee
The length of the vector $\vec{P}$ is 
conserved and is given by its initial value: $P=P_0$.

\item[(2)]
Small $\mu$ regime ~--~ complete late-time de-synchronization. 
At asymptotically large times oscillations completely average out; 
$\vec{P}$ aligns with $\vec{B}$ and 
remains constant, its length having shrunk to the minimal possible value 
$P_{min}=c_{20}P_0$.

\item[(3)]
Intermediate $\mu$ regime -- partial de-synchronization 
at late times. The evolution of $\vec{P}$ at asymptotic times is a precession 
around $\vec{B}$ with some frequency $\omega_s$ (in general, different from 
$\omega_0$), with the length of $\vec{P}$ satisfying $P_{min}<P<P_0$. 
$P_\perp$ 
 remains finite. 
\end{itemize}

\begin{figure}[tbh]
\begin{center}
\includegraphics[height=5.2cm]{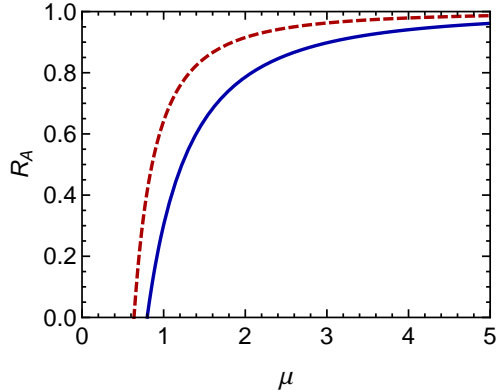}
\end{center}
\caption{\small
Order parameter $R_A\equiv P_\perp/(P_0\sin 2\theta_0)$ at late times for 
Gaussian (solid curve) and box-type (dashed curve) neutrino spectra, 
$\theta_0=\pi/4$.   
Reprinted  with permission from \cite{RaffTamb}; copyright 
2010 by the American Physical Society.
}
\label{fig:RaffTamb}
\end{figure}

These results are illustrated by Fig.~\ref{fig:RaffTamb}, where the late-time 
value of $R_A\equiv P_\perp/(P_0\sin 2\theta_0$) is plotted as a function of 
$\mu$ for Gaussian and box-type neutrino spectra. 
A very interesting feature of the $\mu$-dependence of $R_A$ is its 
threshold behaviour: decoherence is achieved for all values of $\mu$ below a 
certain threshold value $\mu_0$ which depends on the neutrino spectrum and 
is of order of its effective width $\sigma_\omega$. This was not {\it 
a priori} expected -- it could well be that the complete decoherence would 
have occurred only in the limit $\mu\to 0$, with the curves on 
Fig.~\ref{fig:RaffTamb} smoothly approaching the origin. 

In addition to numerical studies, in Ref.~\cite{RaffTamb} also an 
analytical approach was developed, based on the following assumptions 
and approximations: 
\begin{itemize}
\item[(i)] It is assumed that the asymptotic (late time) evolution of the 
global flavour spin vector $\vec{P}$ is a simple precession around $\vec{B}$ 
with a constant frequency $\omega_s$.  

\item[(ii)]
``Sudden approximation'': $\vec{P}(t)$ is replaced by its asymptotic 
 expression starting immediately at $t=0$.

\item[(iii)]
The angles between the individual $\vec{P}_\omega$ and the vectors 
$\vec{B}$ and $\vec{P}$ at the onset of asymptotic regime are taken to be 
those corresponding to $t=0$.

\item[(iv)]
In the corotating frame (the frame rotating around $\vec{B}$ together with 
$\vec{P}$) the individual flavour spin vectors $\vec{P}_\omega'$ 
are replaced by their asymptotic averages assumed to be given by their 
projections on $\vec{H}_\omega'$: $\vec{P}_\omega'\to \langle\vec{P}_\omega'
\rangle= 
\frac{\vec{P}_\omega'\!\cdot\!\vec{H}_\omega'}{H_\omega^{'2}}\vec{H}_\omega'$.
Here the primed quantities refer to the corotating frame.  

\end{itemize}
The analytical results based on the above approximations reproduced very well 
the results of the numerical calculations of $P_\perp$ performed in 
\cite{RaffTamb}. Yet, the underlying assumptions 
were of heuristic nature and are certainly rather far from being realistic. 
Note also that assumptions (ii) and (iii) are not fully consonant: as $\vec{P}$ 
is the sum of all $\vec{P}_\omega$, it does not seem to be consistent 
to take, at the onset of the asymptotic regime, for $\vec{P}_\omega$ their 
values specified by the initial conditions and 
for $\vec{P}$ its asymptotic value (even if the asymptotic regime 
sets in immediately after $t=0$). 
It is also interesting to note that the analytical  
results obtained in \cite{RaffTamb} violate the conservation laws 
(\ref{eq:Etot2}) and (\ref{eq:cons2a}) (except in the limit $\mu\gg\mu_0$, 
where no decoherence occurs). We will discuss 
this point in more detail in Section~\ref{sec:summary}. 
We will also explain there 
the reasons why the analytical approach of \cite{RaffTamb} 
works well despite its underlying assumptions being 
rather unrealistic. In the next subsections we develop two different 
analytical approaches, 
based on our spectral moments formalism augmented only by very simple 
assumptions about the 
late-time behaviour of the global flavour spin vector 
$\vec{P}$ (such as assumption (i) discussed above), without 
invoking any additional conjectures. 

\subsection
{\label{sec:simplified}A simplified analytical approach}

Eqs.~(\ref{eq:Kndiff1})-(\ref{eq:cons2}) and (\ref{eq:cons2a}) describing 
flavour evolution of a dense uniform and isotropic neutrino gas are exact 
and are satisfied for all $t$. We shall now 
 assume that at asymptotically large times the evolution of the system is such 
that the length of the flavour spin vector $\vec{P}$ is conserved. One example 
of such an evolution is the simple precession of $\vec{P}$ around a fixed axis 
in the flavour space, such as the one described in eq.~(\ref{eq:PPP}). 
Thus, our treatment here should be valid if the late-time evolution of the 
system has the form of synchronized collective oscillations, albeit possibly 
with the length of the flavour spin vector $\vec{P}$ decreased by decoherence 
effects. If not noted otherwise, in what follows we will be considering 
all the relevant quantities at asymptotically large times, without 
specifying this explicitly and keeping the same notation for these 
quantities as before.

{}From the evolution equation (\ref{eq:EoM2}) it follows that the conservation 
of $P\equiv|\vec{P}|$ at asymptotically large times implies  
\be
\vec{P}\!\cdot\!\dot{\vec{P}}=\vec{P}\!\cdot\!(\vec{B}\times\vec{S})=0
\,.
\label{eq:PPdot}
\ee
This condition can be satisfied when 
one (or more) of the following conditions is satisfied: 
\mbox{(a) $\vec{P}=0$;} \;(b) $\vec{P}\parallel \vec{B}$; 
\;(c) $\vec{S}=0$; \; (d) $\vec{S}\parallel \vec{B}$; \;(e) $\vec{S}_\perp
\parallel \vec{P}_\perp$. Note that in case (e) only transverse components 
of the vectors $\vec{S}$ and $\vec{P}$ enter, because their components along 
$\vec{B}$ drop out of eq.~(\ref{eq:PPdot}). 
We will not consider case (a) (shrinkage of $\vec{P}$ to zero) which, 
as follows from (\ref{eq:ineq1}),  can only 
be realized when $c_{20}=0$. Cases (b), (c) and (d) are of no interest to us 
either, since they correspond to the situations when at large $t$ not only 
$P\equiv|\vec{P}|$ stays constant, but there is no evolution of $\vec{P}$ 
at all. Therefore we concentrate on case (e), in which at asymptotic times 
$\vec{S}_\perp$ is parallel or antiparallel to $\vec{P}_\perp$, that is 
$\vec{S}_\perp = \omega_s \vec{P}_\perp$. Here we shall make an additional 
assumption that the longitudinal components of $\vec{S}$ and $\vec{P}$ satisfy 
a similar relation with the same proportionality coefficient (we will lift this 
extra assumption in the next subsection). 
In this case we can write 
\be
\vec{S}(t)=\omega_s\vec{P}(t)\,.
\label{eq:S2}
\ee
In principle, the quantity $\omega_s$ could be time dependent; however, 
as we shall show below, eq.~(\ref{eq:Etot2}) implies that it is constant. 

Substituting (\ref{eq:S2}) into (\ref{eq:EoM2}), we find that $\vec{P}$ 
satisfies 
\be
\dot{\vec{P}}=\omega_s \vec{B}\times\vec{P}
\,.
\label{eq:EoM3}
\ee
Thus, in the considered case the evolution of the global flavour spin vector 
$\vec{P}$ at late times is a simple precession around $\vec{B}$, and the 
proportionality coefficient 
$\omega_s$ in eq.~(\ref{eq:S2}) is just the frequency of this precession. 
{}From eq.~(\ref{eq:S2}) and the definitions of $\vec{P}$ and $\vec{S}$ 
it follows that 
\be
\omega_s=\frac
{\int d\omega\omega P_{\omega, i}}{\int d\omega {P}_{\omega, i}}\,, 
\label{eq:omegas}
\ee
where $P_{\omega, i}$ $(i=x,y,z)$
is any of the components of the vector 
$\vec{P}_\omega$. Consistency of our approach requires that the ratio in 
eq.~(\ref{eq:omegas}) be independent of $i$. 

Eq.~(\ref{eq:EoM3}) describes synchronized neutrino oscillations. 
Substituting (\ref{eq:S2}) into the conservation law for ${\cal E}$ given 
in eq.~(\ref{eq:Etot2}) and using eq.~(\ref{eq:PB}), we obtain 
\be
\frac{\mu}{2}[P_0^2-P^2]=c_{20}P_0(\omega_0-\omega_s)
\,. 
\label{eq:Etot3}
\ee

Let us discuss the consequences of 
 this relation. First, we note that all the quantities in it except possibly 
$\omega_s$ are constant, so $\omega_s$ must be constant as well. Second, since 
the left hand side of (\ref{eq:Etot3}) is non-negative, so must be its right 
hand side, i.e. $\omega_s\le \omega_0$. Third, because $\omega_0$ and 
$\omega_s$ are certain averages of $\omega$ over the same spectrum $g_\omega$ 
characterized by an effective width $\sigma_\omega$, their difference cannot 
exceed $\sigma_\omega$ by much, that is 
\be
\omega_0-\omega_s\lesssim \sigma_\omega\,.
\label{eq:omegadiff}
\ee
{}From the latter and eq.~(\ref{eq:Etot3}) it immediately follows that in the 
limit $\mu\to \infty$ we must have $P\to P_0$ (no shrinkage of $\vec{P}$ 
for very large $\mu$). 
Later on we will also demonstrate that in the limit $\mu\to\infty$ the 
frequency of synchronized oscillations $\omega_s$ approaches $\omega_0$. 

Eq.~(\ref{eq:Etot3}) also allows us to clarify the meaning of the formal limit 
$\mu\to\infty$. In practical terms, this limit means that $\mu$ becomes large 
compared to some quantity of dimension of energy characterizing the neutrino 
system under consideration. In our case this could be e.g.\ $\omega_0$, 
$\omega_s$ or $\sigma_\omega$. We shall now show that this characteristic 
parameter is actually $\sigma_\omega$. 
Indeed, let us rewrite (\ref{eq:Etot3}) as 
\be
\frac{P_0^2-P^2}{P_0^2}\,=\,2c_{20}\frac{\omega_0-\omega_s}{\mu P_0}
\,\lesssim \, 2c_{20}\,\frac{\sigma_\omega}{\mu P_0}\,, 
\label{eq:Etot3a}
\ee
where the approximate inequality follows from (\ref{eq:omegadiff}). From 
(\ref{eq:Etot3a}) 
one can immediately see that `perfect synchronization', when the 
asymptotic value of $P$ coincides with its initial value $P_0$ 
and decoherence effects are negligible, is achieved for 
$\mu P_0\gg \sigma_\omega$. At the same time it is irrelevant 
whether or not $\mu P_0$ is large compared to $\omega_0$ (or to $\omega_s$). 
This is in accord with the fact that $\omega_0$ or $\omega_s$ can always 
be eliminated by going into a proper rotating frame.
Thus, one can expect noticeable decoherence effects only for 
$\mu P_0\lesssim \sigma_\omega$.

Eq.~(\ref{eq:Etot3}) relates two unknowns -- the late-times value of 
$P$ and the frequency of synchronized oscillations $\omega_s$. To find these 
quantities, we need one more relation between them. As such, we will use the 
new conservation law (\ref{eq:cons2}) that was derived in 
Section~\ref{sec:specmoments}. Let us first note that 
eqs.~(\ref{eq:S2}) and (\ref{eq:EoM3}) together with time independence of 
$\omega_s$ imply that at asymptotic times the vector $\vec{S}$ satisfies the 
EoM similar to (\ref{eq:EoM3}), 
\be
\dot{\vec{S}}=\omega_s\vec{B}\times\vec{S}\,,
\label{eq:EoM4}
\ee
and so its length is conserved: $\vec{S}\!\cdot\!\dot{\vec{S}}=0$. On the other 
hand, from eq.~(\ref{eq:Kndiff1}) with $n=1$ we have an exact relation 
\be
\dot{\vec{S}}=\vec{B}\times\vec{K}_2 +\mu\vec{P}\times\vec{S}\,.
\label{eq:EoM5}
\ee
The conservation of the length of $\vec{S}$ then implies 
\be
\vec{S}\!\cdot\!(\vec{B}\times\vec{K}_2)=0\,.
\label{eq:SSdot}
\ee
Following now the arguments similar to those given just below 
eq.~(\ref{eq:PPdot}), we conclude that eq.~(\ref{eq:SSdot}) is nontrivially 
realized only if at asymptotically large times $\vec{K}_2$ is parallel or 
antiparallel to $\vec{S}$ (and therefore also to $\vec{P}$), that is 
$\vec{K}_2(t) = \omega_1 \vec{S}(t)$. Comparing then eqs.~(\ref{eq:EoM5}) 
and~(\ref{eq:EoM4}), we find $\omega_1=\omega_s$, that is, asymptotically,   
\be
\vec{K}_2(t) = \omega_s \vec{S}(t)=\omega_s^2 \vec{P}(t)\,.
\label{eq:K2}
\ee  
Substituting this into (\ref{eq:cons2a}) yields 
\be
-\omega_s^2c_{20}P_0+\omega_s\mu P^2=P_0[\mu P_0\omega_0-
c_{20}(\omega_0^2+\sigma_\omega^2)]\,.
\label{eq:cons2b}
\ee
Excluding now $P^2$ from eqs.~(\ref{eq:Etot3}) and (\ref{eq:cons2b}), 
we find the following 
equation for 
$\omega_0-\omega_s$:
\be
(\omega_0-\omega_s)^2-\frac{\mu P_0}{c_{20}}(\omega_0-\omega_s)+\sigma_\omega^2
=0\,.
\label{eq:quadr}
\ee
Its solution is 
\be
\omega_0-\omega_s=\frac{1}{2c_{20}}\Big[\mu P_0-
\sqrt{\mu^2 P_0^2-4c_{20}^2\sigma_\omega^2}\,\Big].
\label{eq:sol1}
\ee
We have discarded the solution with the plus sign in front of the square root 
because for $\mu P_0\gg \sigma_\omega$ it would lead to $\omega_0-\omega_s
\gg \sigma_\omega$, in contradiction with (\ref{eq:omegadiff}). 

Let us now discuss eq.~(\ref{eq:sol1}). First, we notice that it exhibits a 
threshold behaviour: the real solution only exists (and therefore synchronized 
oscillations can only take place) for $\mu > \mu_0$, where $\mu_0$ 
is given by 
\be
\mu_0 P_0 = 2c_{20}\sigma_\omega
\,.
\label{eq:mu0}
\ee 
Next, we find that for $\mu P_0 \gg 2c_{20}\sigma_\omega$ (i.e.\ far above 
the threshold) 
\be
\omega_0-\omega_s\simeq c_{20}\frac{\sigma_\omega^2}{\mu P_0}
\ll \sigma_{\omega}\,. 
\label{eq:ineq2}
\ee
Thus, for large $\mu P_0$ the precession frequency $\omega_s \to \omega_0$. 
The difference $\omega_0-\omega_s$ reaches its maximum at the threshold 
$\mu=\mu_0$:
\be
\omega_0-\omega_s(\mu_0)=\sigma_\omega\,.
\label{eq:diff1}
\ee

By using $\omega_0-\omega_s$ from eq.~(\ref{eq:sol1}) in eq.~(\ref{eq:Etot3}) 
(or equivalently in eq.~(\ref{eq:cons2b})), one can find the asymptotic value 
of $P$ as a function of $\mu$:
\be
P=P_0\Big(1-\frac{\mu_0^2}{\mu^2}\Big)^{1/4}.
\label{eq:Pas1}
\ee

The above results for the asymptotic regime 
have several attractive features: 
they demonstrate in a very simple way the existence of the 
threshold $\mu_0$ below which no synchronized oscillations are possible, give 
the correct order-of-magnitude estimate for its value ($\mu_0 P_0 \sim 
\sigma_\omega$), and lead to a reasonable behaviour of $P$ far above the 
threshold. However, 
eq.~(\ref{eq:Pas1}) gives a wrong 
value of $P$ at the 
threshold: $P=0$. This is obviously incorrect because, due to the conservation 
of $\vec{P}\!\cdot\!\vec{B}=-c_{20}P_0$, the length of $\vec{P}$ cannot be 
smaller than $c_{20}P_0$. 
A possible reason for this is that eq.~(\ref{eq:S2}) involves an additional  
assumption about the longitudinal components of $\vec{P}$ and $\vec{S}$ 
which actually does not directly follow from eq.~(\ref{eq:PPdot}). 
It can be shown that this additional 
assumption (and so the relation in eq.~(\ref{eq:S2})) is actually well 
satisfied far above the threshold $\mu_0$ but breaks down 
close to the threshold. 

Indeed, we have found that for $\mu\gg \mu_0$ the asymptotic length of 
$\vec{P}$ coincides with $P_0$. This means that the time interval between 
$t=0$ and the onset of the asymptotic regime, during which $P$ may shrink, 
is essentially zero, and the $P$-preserving synchronized oscillations 
described by eq.~(\ref{eq:EoM3}) 
start practically immediately at $t=0$. Thus, $P$ is conserved at all times. 
{}From eq.~(\ref{eq:Etot1}) it then follows 
that $\vec{B}\!\cdot\!\vec{S}$ is also conserved at all times.
Together with the relation $\vec{S}_\perp(t)=\omega_s\vec{P}_\perp(t)$ 
this means that, starting practically from 
$t=0$, the vector $\vec{S}$ precesses around $\vec{B}$ with the angular 
velocity $\omega_s$, just as $\vec{P}$ does.   
Since $\vec{S}$ and $\vec{P}$ are collinear at $t=0$ (both pointing in the 
$z$-direction in the flavour space), they will then remain collinear at all 
times. Obviously, this argument fails close to the threshold $\mu_0$. 

In the next subsection we shall lift the additional assumption 
about the longitudinal components of $\vec{S}$ and $\vec{P}$. 
As we shall see, the assumption about the asymptotic behaviour of $P_\perp$ 
will also need to be corrected. 

\subsection{\label{sec:better} 
An analytical approach valid for all $\mu$} 

We shall now attempt to develop an approximate analytical approach valid for 
all $\mu$.   

\subsubsection{\label{sec:const}Approximation of constant asymptotic value 
of $P$}

Let us first assume, as we did in the Section~\ref{sec:simplified}, that 
the evolution of the system at asymptotically large times conserves the length 
of the vector $\vec{P}$, that is, eq.~(\ref{eq:PPdot}) is satisfied. As was 
discussed above, a nontrivial realization of this condition requires the 
following relation between the transverse components of $\vec{P}$ and 
$\vec{S}$ at 
late times: 
\be
\vec{S}_\perp(t)=\omega_s\vec{P}_\perp(t)
\,.
\label{eq:S3}
\ee
Unlike we did in Section~\ref{sec:simplified}, we will not make here the 
additional assumption $\vec{S}_\parallel=\omega_s \vec{P}_\parallel$, which 
is valid only far above the threshold $\mu_0$.

Substituting eq.~(\ref{eq:S3}) into eq.~(\ref{eq:EoM2}), 
we again obtain the EoM (\ref{eq:EoM3}) describing the precession of the 
flavour spin vector $\vec{P}$ around $\vec{B}$ with the angular velocity 
$\omega_s$. In our analysis in Section~\ref{sec:simplified}  
we found that the parameter $\omega_s$ introduced through eq.~(\ref{eq:S2}) 
was time-independent; this followed from the conservation law 
(\ref{eq:Etot1}) (along with $\vec{B}\!\cdot\!\vec{P}=const.$ and the 
late-time relation $P=const$). Here we cannot use the same argument, as 
the parameter $\omega_s$ now relates only the transverse components of the 
vectors $\vec{P}$ and $\vec{S}$, whereas the conservation law (\ref{eq:Etot1}) 
constrains only the longitudinal component of $\vec{S}$. 
Instead, we will {\it assume} 
that the parameter $\omega_s$ can be considered as constant at asymptotically 
large times.
Thus, our assumption here about the late-time behaviour of the flavour spin 
vector in fact coincides with assumption (i) discussed in 
Section~\ref{sec:previous}. However, we will {\sl not} use listed 
there assumptions (ii) - (iv), employed in \cite{RaffTamb}; our spectral 
moments formalism will allow us to fully determine the late-time behaviour of 
the system basing solely on assumption (i). 

It is easy to demonstrate by induction that at asymptotic times all the 
spectral moments $\vec{K}_n$ satisfy the evolution equations similar to 
(\ref{eq:EoM3}) 
with all 
$\vec{K}_{n\perp}$ being collinear with $\vec{P}_\perp$: 
\be
\dot{\vec{K}}_n=\omega_s \vec{B}\times\vec{K}_n\,,\qquad \quad 
\vec{K}_{n\perp}=\alpha_n\vec{P}_\perp\,,
\label{eq:EoM7}
\ee
where $\alpha_n$ is constant. Similar relations hold also for the flavour spin 
vectors of the individual modes $\vec{P}_\omega$. The derivation is especially 
simple in the corotating frame, see the Appendix. 
As shown there, for $P_\perp\ne 0$  
one can then find the ratio of the longitudinal and transverse 
components of the asymptotic vectors $\vec{P}_\omega$. 
Combining this with the normalization condition $P_{\omega\parallel}^2
+P_{\omega\perp}^2=P_\omega^2=P_0^2g_\omega^2$ allows one to determine 
$P_{\omega\parallel}$ and $P_{\omega\perp}$. 
With the expressions for these quantities at hand, the longitudinal and 
transverse components of all the spectral moments $\vec{K}_n$ at asymptotic 
times can then be found  from eq.~({\ref{eq:Kn}). 
\begin{figure}[t]
\begin{center}
\includegraphics[height=6.2cm]{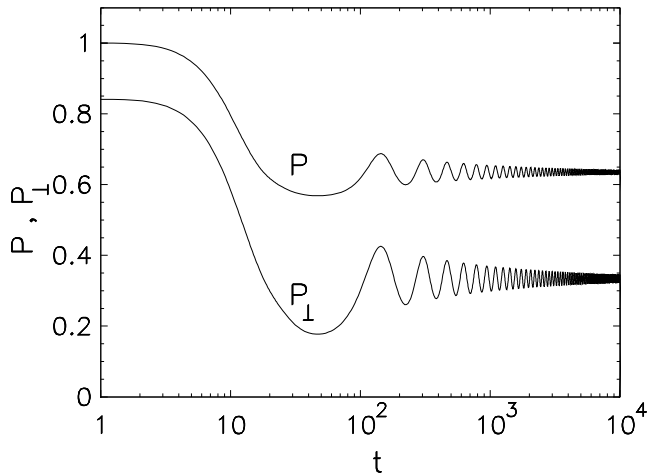}
\vspace*{-0.8cm}
\end{center}
\caption{\small Time dependence of $P$ (upper curve) and 
$P_\perp$ (lower curve). Gaussian neutrino spectrum,
$P_0=1$, $\omega_0=1$, $\sigma_\omega=0.1$, $\theta_0=0.5$,  
$\mu=0.108$. 
}
\label{fig:Plong} 
\end{figure} 
However, the direct calculation shows that, 
except in the limit $\mu\to\infty$, the values of $P_\perp$ obtained in this 
way do not reproduce correctly the results of numerical integration of the 
exact EoMs; in particular, no threshold behaviour in $\mu$ is found. This 
means that the assumption that $P\equiv|\vec{P}|$ approaches a constant value 
at asymptotically large times, on which our consideration here was thus far 
based, must actually be incorrect. 

This point is partly illustrated by Fig.~\ref{fig:Plong}, which shows that 
for $\mu$ exceeding the critical value $\mu_0$ (but not far above it) the 
quantities $P_\perp$ and $P$ keep oscillating around their mean values even 
at very late times. Obviously, this 
figure cannot serve as a proof that the oscillations will survive for all 
times (which actually follows from our analytical results), but it shows 
that these oscillations continue even up to $t=10000$ (in the units in which 
$\omega_0=1$), which is several orders of magnitude larger than the naively 
expected coherence length $L_{\rm coh}\simeq 1/(2\sigma_\omega)$. The 
situation is different for $\mu\gg \mu_0$, when $P$ is practically constant 
at all times.

\subsubsection{\label{sec:aver}Invoking the averaging procedure}

As the amplitude of the late-time oscillations of $P_\perp$ is 
relatively small, it seems to be a reasonable approximation 
to replace it at asymptotic times by its mean value, found by averaging over 
these oscillations. 
However, as our analysis in the previous subsection shows, this should be done 
through a consistent averaging procedure rather than by simply assuming that at 
large times $P_\perp=const$. 
To carry out the averaging systematically, we average the EoMs of the 
flavour spin vectors over a very large (formally infinite) time interval. 
It is convenient to do this 
 in the corotating frame, where the late-time precession of the flavour spin 
vector $\vec{P}$ is `rotated away' from its evolution,  and 
its direction at late times is essentially fixed (see the Appendix). 
Note that going to the corotating frame does not change the longitudinal 
(with respect to $\vec{B}$) components of the flavour spin vectors and 
spectral moments as well as the lengths of their transverse components.

The relevance of the averaging procedure for studying the behaviour 
of the system at late times 
follows from the well known property of infinite-interval averages, 
that for any function $f(t)$ that is integrable on any finite interval in 
$[0,\infty)$, the average can be found as 
\be
\langle f(t)\rangle=\lim\limits_{t\to\infty}f(t)\,
\label{eq:av1}
\ee
provided that the limit on the right hand side exists. If at $t\to\infty$ the 
limit in (\ref{eq:av1}) does not exist but $f(t)\to f_0$ + oscillating terms, 
where $f_0$ is a finite constant and the oscillating terms are of finite 
amplitude and average to zero, then $\langle f(t)\rangle=f_0$.  

As demonstrated in the Appendix, the averaging procedure allows us to find 
the ratio of the averages of the transverse and longitudinal components of 
$\vec{P}_\omega$, 
which can be cast into the form 
\be
\langle \vec{P}_{\omega\perp}\rangle=f(\omega,\mu)
\langle\mu\vec{P}_\perp\rangle\,,\qquad~
\nonumber 
\ee
\be
\langle P_{\omega\parallel}\rangle=f(\omega,\mu)
\big[\omega-\omega_r(\mu)\big]\,,
\label{eq:Pomegas}
\ee
with as yet unknown function $f(\omega,\mu)$. 
Note that we cannot combine 
the ratio $\langle P_{\omega\perp}\rangle/\langle 
P_{\omega\parallel}\rangle$ with the normalization 
condition for $P_\omega^{2}$ in order to find $\langle P_{\omega\perp}
\rangle$ and $\langle P_{\omega\parallel}\rangle$ separately  
since these averages do not satisfy the same normalization condition as 
the un-averaged quantities $P_{\omega\perp}$ and $P_{\omega\parallel}$. 
Therefore, in order to determine 
$\langle P_{\omega\parallel}\rangle$ and $\langle\vec{P}_{\omega\perp}
\rangle$ one needs one more relation between them. 
As shown in the Appendix, such a relation can be found by considering the 
average $\langle\vec{P}_\omega\!\cdot\!\vec{H}_\omega\rangle$ in the 
corotating frame. 
Together with eq.~(\ref{eq:Pomegas}), this gives 
\be
f(\omega,\mu)=P_0 g_\omega\frac{
-c_{20}(\omega-\omega_r)+s_{20}\langle\mu P_\perp\rangle}
{(\omega-\omega_r)^2+\langle\mu P_\perp\rangle^2}\,.
\label{eq:f1}
\ee
Here we have introduced the notation
\be
\omega_r
\equiv \omega_s(\mu)+c_{20}\mu P_0\,.
\label{eq:omegar}
\ee
Next, we calculate the averages%
\footnote{Note  
that $\langle \vec{P}_{\omega\perp}\rangle$ and 
$\langle\mu\vec{P}_\perp\rangle$ are parallel, which follows from the first 
equation in~(\ref{eq:Pomegas}).
}
\be
\langle P_\perp\rangle =
\int d\omega \langle P_{\omega\perp}\rangle\,,\qquad 
\langle P_\parallel\rangle =
\int d\omega \langle P_{\omega\parallel}\rangle\,.
\label{eq:relations}
\ee
{}From eqs.~(\ref{eq:Pomegas}), (\ref{eq:f1}) and (\ref{eq:PB}) it follows 
that these relations can be rewritten as 
\be
1=\mu P_0\int d\omega 
g_\omega\frac{
-c_{20}(\omega-\omega_r)+s_{20}\langle\mu P_\perp
\rangle}{(\omega-\omega_r)^2+\langle\mu P_\perp\rangle^2}\,,\qquad\quad
\label{eq:rel1}
\ee
\be
-c_{20}=
\int d\omega g_\omega\frac{
-c_{20}(\omega-\omega_r)+s_{20}\langle\mu P_\perp\rangle}
{(\omega-\omega_r)^2+\langle\mu P_\perp\rangle^2}(\omega-\omega_r)\,. 
\vspace*{1.5mm}
\label{eq:rel2}
\ee
For any neutrino spectrum function $g_\omega$ and a given $\mu$, these two 
equations can be solved for the two unknowns, $\omega_s(\mu)$ and 
$\langle P_\perp(\mu)\rangle$. Note that eq.~(\ref{eq:rel1}) has been 
obtained by dividing both sides of the first equality in (\ref{eq:relations}) 
by $\langle P_\perp\rangle$, and so it is a necessary condition for the 
existence of a nontrivial solution $\langle P_\perp\rangle\ne 0$. 

Eqs.~(\ref{eq:rel1}) and (\ref{eq:rel2}) coincide with eq.~(17) of 
\cite{RaffTamb} (notice that our definition of the sign of $c_{20}$ is opposite 
to theirs). However, we did not use the `sudden approximation' and the 
assumption that at the onset of the asymptotic regime the flavour spin vectors 
of the individual modes $\vec{P}_\omega$ have the same directions as they have 
at $t=0$, which were employed in \cite{RaffTamb}. 
As was pointed out in Section~\ref{sec:previous}, these assumptions are in 
general not well justified. Our consideration was instead based on the 
averaging procedure performed at the level of EoMs and the assumption that at 
asymptotically large times $P_\perp=|\vec{P}_\perp|$ undergoes only relatively 
small oscillations (though it is not constant). The latter is confirmed by 
our numerical calculations.

After a little algebra one can obtain from eqs.~(\ref{eq:rel1}) and 
(\ref{eq:rel2}) a simpler pair of equations \cite{RaffTamb}:
\be
\frac{s_{20}}{\mu}=P_0\int d\omega g_\omega
\frac{\kappa}{(\omega-\omega_r)^2+\kappa^2}\,,
\label{eq:RT19a}
\ee
\be
-\frac{c_{20}}{\mu}=
P_0\int d\omega g_\omega\frac{(\omega-\omega_r)}{(\omega-\omega_r)^2+
\kappa^2}\,.\;\,
\label{eq:RT19b}
\ee
Here the notation $\kappa\equiv \langle\mu P_\perp\rangle$ has been introduced. 
Noting that at the limit $\mu\to\mu_0$ one has $\kappa\to 0$ and the integrand 
of (\ref{eq:RT19a}) goes to $g_\omega\pi\delta[\omega-\omega_r(\mu_0)]$, 
one finds that in this limit eqs.~(\ref{eq:RT19a}) and (\ref{eq:RT19b}) become
\be
\frac{s_{20}}{\mu_0}=P_0\pi g_{\omega_r(\mu_0)}\,,\qquad\qquad
-\frac{c_{20}}{\mu_0}=
{\cal P}\int d\omega g_\omega\frac{P_0}{\omega-\omega_r(\mu_0)}\,,
\label{eq:mu0RT}
\ee
where 
${\cal P}$ stands for the Cauchy principal value. These equations can be 
solved to find the threshold value $\mu_0$ and $\omega_r(\mu_0)$ 
\cite{RaffTamb}. 

As was mentioned above, given the neutrino spectrum $g_\omega$, 
eqs.~(\ref{eq:rel1}) and (\ref{eq:rel2}) (or equivalently (\ref{eq:RT19a}) and 
(\ref{eq:RT19b})) can be solved numerically for $\omega_s$ and $\kappa=
\langle\mu P_\perp\rangle$. We will, however, consider now the simple box-type 
spectrum 
\be
g_\omega=\frac{1}{2\sigma}\cdot\left\{\begin{array}{ll}
1, ~|\omega-\omega_0|\le \sigma,\\
0, ~|\omega-\omega_0|> \sigma
\end{array} \right.,
\label{eq:box}
\ee
for which the explicit analytical solution of the problem can be found. 
Note that the parameter $\sigma$ here is related to the variance 
$\sigma_\omega^2\equiv \langle\omega^2\rangle-\langle\omega\rangle^2$ as 
$\sigma_\omega^2=\frac{1}{3}\sigma^2$. As the $\omega$-spectrum (\ref{eq:box}) 
is flat, the threshold value $\mu_0$ can be immediately found from the 
first equation in~(\ref{eq:mu0RT}): 
\be
\mu_0 P_0=\frac{2}{\pi}\,s_{20}\,\sigma 
\,.
\label{eq:mu0RTa}
\ee
For $\mu \ge \mu_0$ eqs.~(\ref{eq:RT19a}) and (\ref{eq:RT19b}) yield 
\be
s_{20}\frac{2\sigma}{\mu P_0}=
\arctan\Big(
\frac{\omega_0-\omega_r+\sigma}{\kappa}\Big) -\arctan\Big(
\frac{\omega_0-\omega_r-\sigma}{\kappa}\Big)\,,
\label{eq:RTtrans1}
\ee 
\be
-c_{20}\frac{4\sigma}{\mu P_0}=
\ln\frac{(\omega_0-\omega_r+\sigma)^2+\kappa^2}
{(\omega_0-\omega_r-\sigma)^2+\kappa^2}\,.
\label{eq:RTtrans2}
\ee 
For the consistency of the last relation it is necessary that $(\omega_0-
\omega_r)$ be negative. 
The pair of transcendental equations~(\ref{eq:RTtrans1}) and 
(\ref{eq:RTtrans2}) admits analytical solution 
for the two unknowns, $\kappa$ and $\omega_0-\omega_r$. 
Noting that $\langle P_\perp\rangle=\kappa/\mu$, we obtain  
\be
\langle P_\perp\rangle=
\frac{2\sigma}{\mu}\!\cdot\!\frac{\sin(s_{20}\frac{2\sigma}{\mu P_0})}
{\exp{(c_{20}\frac{2\sigma}{\mu P_0})}
+\exp(-c_{20}\frac{2\sigma}{\mu P_0})-
2\cos(s_{20}\frac{2\sigma}{\mu P_0})}\,, 
\label{eq:PperpRT}
\ee
\be
\omega_0-\omega_r=
-\sigma\!\cdot\!\frac{\exp{(c_{20}\frac{2\sigma}{\mu P_0})}
-\exp(-c_{20}\frac{2\sigma}{\mu P_0})}
{\exp{(c_{20}\frac{2\sigma}{\mu P_0})}
+\exp(-c_{20}\frac{2\sigma}{\mu P_0})-
2\cos(s_{20}\frac{2\sigma}{\mu P_0})}\,.
\vspace*{1.5mm}
\label{eq:omegaDiff}
\ee
The asymptotic precession frequency $\omega_s(\mu)$ can then be found from 
eqs.~(\ref{eq:omegaDiff}) and (\ref{eq:omegar}). 

In Fig.~\ref{fig:comparison} we compare the results of direct numerical 
integration of EoMs of the flavour spin vectors with the results of the 
approximate analytical approach described here. 
In the left panel
we plot the order parameter $R_A=\langle P_\perp\rangle/P_\perp(0)=\langle 
P_\perp\rangle/(s_{20}P_0)$, which describes decoherence effects, as a 
function of $\mu$. The results are presented for two neutrino spectra: the 
Gaussian spectrum (\ref{eq:Gauss}) and the box-type spectrum of 
eq.~(\ref{eq:box}). The right panel of Fig.~\ref{fig:comparison} presents a 
similar comparison for the asymptotic precession frequency $\omega_s$. 

For the Gaussian spectrum the numerical solution of EoMs was obtained 
considering $N_{\omega}=3000$ modes uniformly distributed in the interval  
$(\omega_0-4 \sigma_{\omega},\, \omega_0+ 4 \sigma_{\omega})$. The same 
number of modes was used for the box-type spectrum. We 
considered the flavor evolution in the time range $t\in [0,\,10000]$. 
The error due to the discretization of the neutrino spectrum was estimated 
by doubling the number of the modes and is completely negligible 
($\sim 10^{-5}$). For the average value of $P_\perp$ at late times we took 
the arithmetic mean of the last maximum and last minimum of the oscillating 
curve before $t=10000$ (see Fig.~\ref{fig:Plong}). The quantity $\omega_s$ 
was found numerically from eq.~(\ref{eq:S3}).  

For the approach based on the averaging procedure, 
in the case of the Gaussian spectrum we solve eqs.~(\ref{eq:RT19a}) and 
(\ref{eq:RT19b}) numerically, whereas for the box-type spectrum the fully 
analytical solution in eqs.~(\ref{eq:PperpRT}) and (\ref{eq:omegaDiff}) is 
used. Note that near coincidence of the $R_A$ curves corresponding to 
the two spectra that we used is a curious accident of our choice of the value 
of $\theta_0$: we have checked that for other choices the curves are clearly 
distinguishable, though similar in shape. It can be seen from the figure that
the analytical approach reproduces the results of numerical integration 
of the exact EoMs extremely well. 
This is in accord with the conclusions of Ref.~\cite{RaffTamb}, where 
eqs.~(\ref{eq:RT19a}) and (\ref{eq:RT19b}) were first obtained. 

\begin{figure}[tbh]
\vspace*{0.6cm}
\begin{center}
\includegraphics[height=5.0cm]{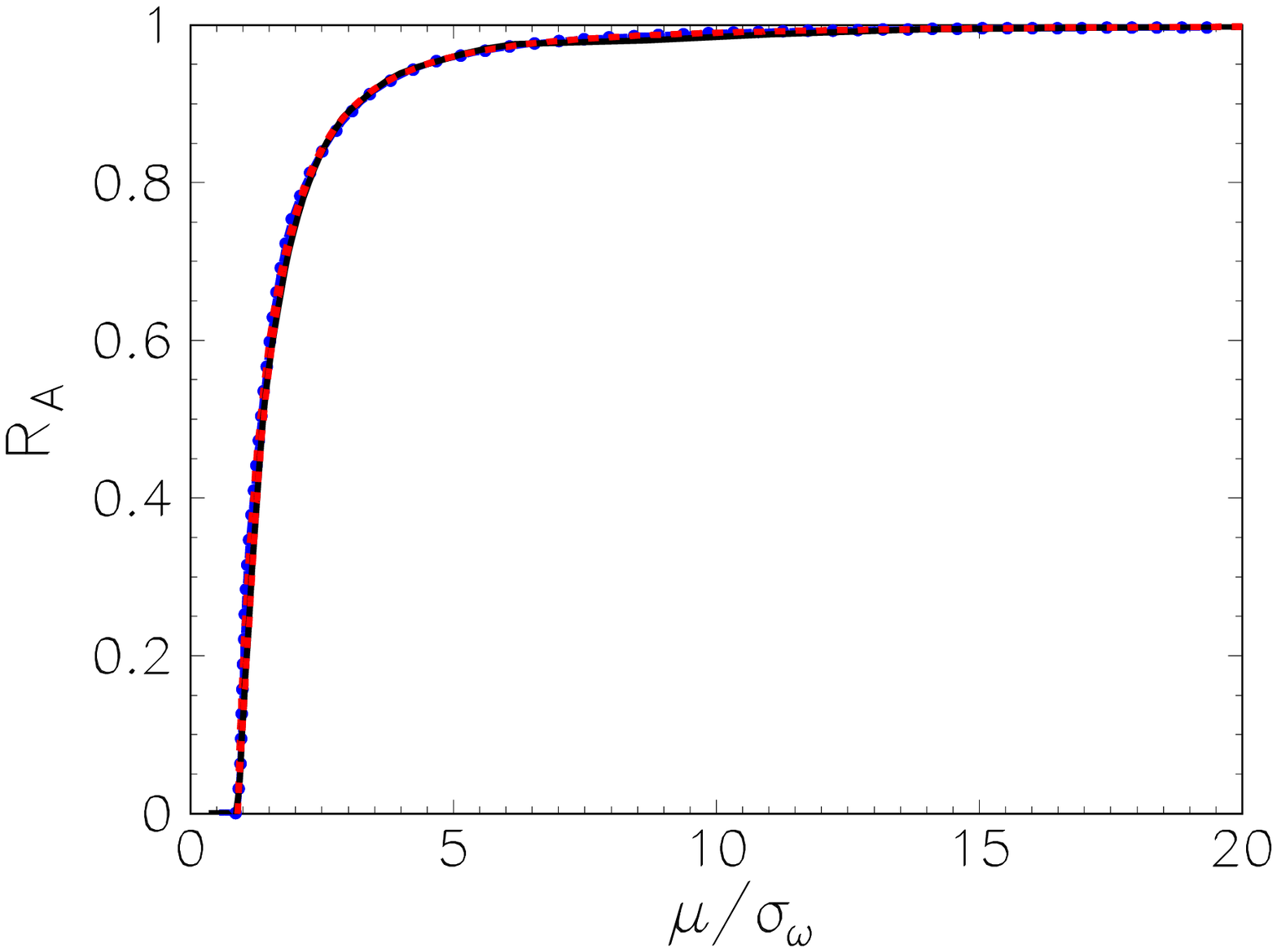}
\hspace*{1.0cm}
\includegraphics[height=5.0cm]{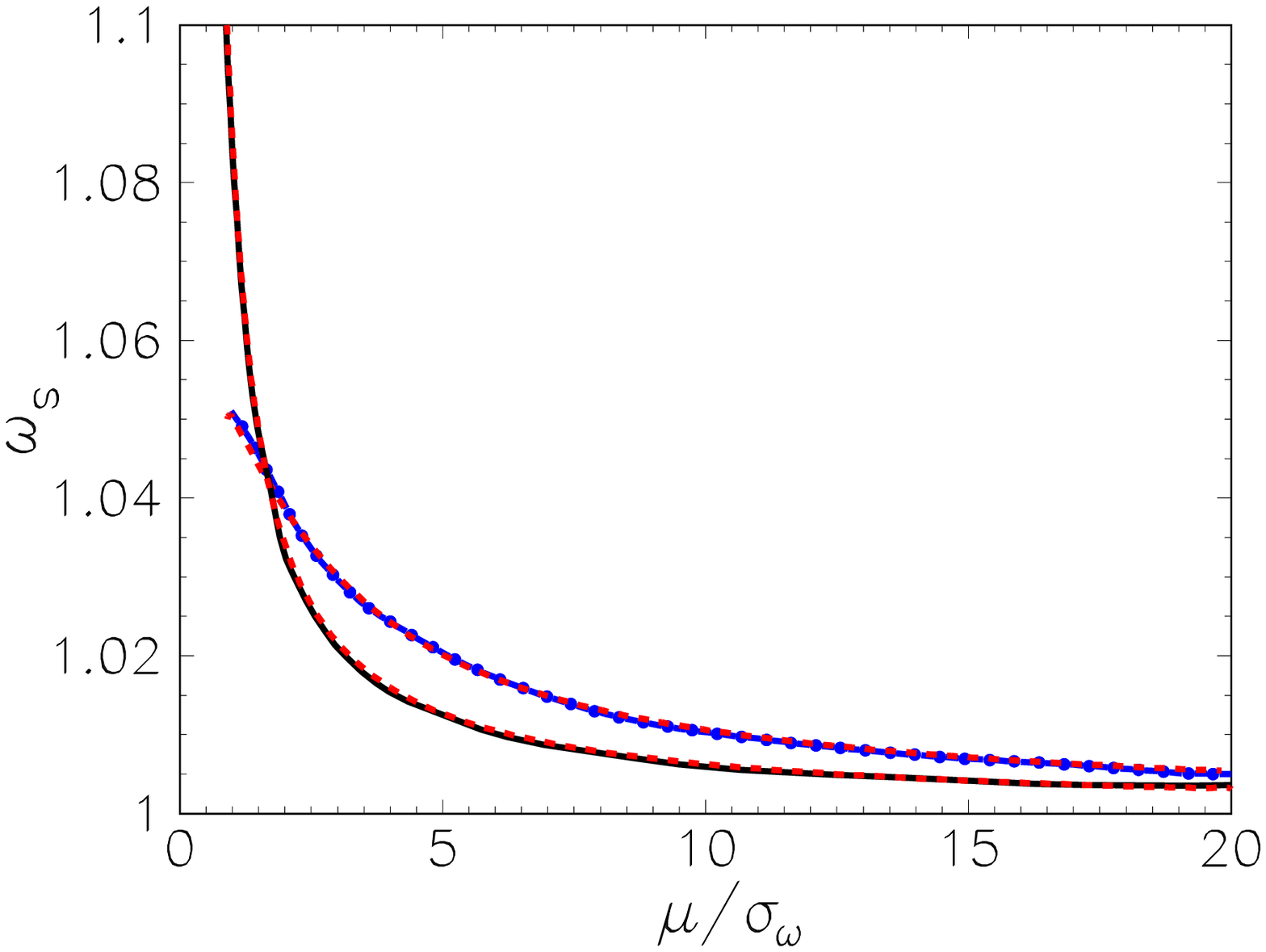}
\end{center}
\caption{\small Left panel: Order parameter $R_A\equiv\langle P_\perp\rangle/
(P_0\sin 2\theta_0)$. Dash-dotted (blue) curve: numerical integration of EoMs 
for Gaussian spectrum, solid (black) curve: same for box-type spectrum. 
Dashed (red) curves: results of the analytical approach for these spectra  
(see the text). For both spectra $P_0=1$, $\omega_0=1$, $\theta_0=0.5$. 
For Gaussian spectrum we use $\sigma_\omega=0.2$, for box-type spectrum  
$\sigma=0.2$, which corresponds to $\sigma_\omega=\sigma/\sqrt{3}\simeq 0.115$. 
Note that near coincidence of $R_A$ curves corresponding to the two 
considered spectra is an accident of our choice of $\theta_0$. 
Right panel: same as left one, but for $\omega_s$. 
}
\label{fig:comparison}
\end{figure}

\section{\label{sec:interp}Decoherence by wave packet separation: 
Adiabaticity and adiabaticity violation} 

Let us try to understand the obtained above results on coherence and 
decoherence of 
the oscillations in a dense neutrino gas from the viewpoint of wave packet 
separation. The split-up of wave packets in configuration space occurs as a 
consequence of the difference of group velocities of the wave packets of 
different neutrino states composing the initially produced neutrino flavour 
eigenstate. Since only the states that diagonalize the Hamiltonian of the 
neutrino system (propagation eigenstates) have well-defined group velocities, 
our discussion will be in terms of these eigenstates.%
\footnote{Collective neutrino oscillations have been previously considered 
from the viewpoint of the propagation eigenstate basis in 
\cite{cristina1,cristina2}, but the issue of wave packets an their 
separation has not been addressed there.
}

Why doesn't decoherence by wave packet separation occur in very dense neutrino 
gases, when the values of the neutrino self-interaction parameter $\mu$ 
are far above the threshold $\mu_0$? One might suspect that 
in this case the dynamics of neutrino evolution is such that the difference 
of the group velocities of different propagation eigenstates $\Delta v_g$ 
vanishes and they all propagate with the same speed. It is, however, easy to 
make sure that this is not the case, and $\Delta v_g$ does {\sl not} vanish. 
As we shall see, it is 
nevertheless possible to qualitatively understand the absence of decoherence 
in the large $\mu$ limit as well as partial decoherence for $\mu$ only 
slightly exceeding 
$\mu_0$ in terms of the behaviour of the neutrino propagation eigenstates. 

For neutrino systems with time-dependent Hamiltonians, such as the one 
described by the Hamiltonian vector~(\ref{eq:vecH2}), the propagation 
eigenstates cannot be defined universally, i.e.\ in a time-independent way. 
However, at any instant of time $t$ they can be defined as the states 
diagonalizing the Hamiltonian at this particular time. These are the so-called 
instantaneous eigenstates of the Hamiltonian. 
At a given time $t$ the neutrino flavour eigenstates can be written as linear 
superpositions of the instantaneous propagation eigenstates. 
As in ordinary matter, these superpositions are determined in the 2-flavour 
case by the mixing angle in matter $\theta=\theta(t)$. 

The propagation eigenstates evolve independently in the adiabatic limit, 
i.e.\ when the neutrino Hamiltonian changes relatively slowly, so that the 
system has enough time to `adjust' itself to the changing conditions. 
In terms of the flavour spin vectors, 
adiabaticity means that the rate of evolution of $\vec{H}_\omega$ 
is small compared to the frequency $H_\omega$ of precession of the individual 
$\vec{P}_\omega$ around their $\vec{H}_\omega$. In this case, in the course of 
their precession around $\vec{H}_\omega$, the vectors $\vec{P}_\omega$  
`track' the movements of $\vec{H}_\omega$. 
If the adiabaticity is violated, the propagation eigenstates do not evolve 
independently; instead, they can go into each other during the 
evolution of the neutrino system. 
As a measure of adiabaticity violation 
one can choose the ratio $\gamma$ of the off-diagonal element of the 
Hamiltonian of the system in the propagation eigenstate basis to the 
difference of its diagonal elements. 
Adiabatic regime corresponds to $\gamma \ll 1$, whereas $\gamma \gg 1$ would 
mean maximal violation of adiabaticity. 
It should be noted that, while for the oscillations in ordinary matter 
adiabaticity may only be violated in the case of non-uniform matter, in dense 
neutrino environments adiabaticity violation may occur even in the case of 
constant neutrino density. 
This happens because in the latter case the Hamiltonian 
of the system depends not only on the overall density of the neutrino gas, but 
also on its flavour composition, which changes with time. 

Consider now several regimes of the evolution of the system, depending on the 
values of the parameter $\mu$. 
\begin{itemize}
\item[I.]
$\mu P_0 \gg \omega_0$. In this case the Hamiltonian vector $\vec{H}_\omega
=\omega\vec{B}+\mu\vec{P}$ nearly coincides with $\mu\vec{P}$. Since at $t=0$ 
all the individual flavour spin vectors $\vec{P}_\omega$ as well as 
their sum $\vec{P}$ point in the 
$z$-direction, 
all $\vec{P}_\omega$ are
practically collinear with their $\vec{H}_\omega$. From EoM (\ref{eq:EoM1}) it 
then follows that there is essentially no flavour evolution in 
this case. 

As the Hamiltonian of the system remains practically constant, the 
adiabaticity 
condition is very well satisfied. The propagation eigenstates are 
well defined and evolve independently. 
The condition $\mu P_0 \gg \omega_0$ means that the mixing is strongly 
suppressed in the neutrino gas, so that the initially produced
flavour state practically coincides with one of the propagation 
eigenstates rather than being a nontrivial superposition of different 
eigenstates. Therefore, no wave packet separation occurs (a propagation 
eigenstate cannot `separate with itself'),  
and hence there is no decoherence.

\item[II.]
$\sigma_\omega\ll \mu P_0 \lesssim \omega_0$. This case is most simply 
considered in the corotating frame, in which one has to replace 
$\omega \to \omega'=(\omega-\omega_s)$. Since $|\omega-\omega_s|\lesssim 
\sigma_\omega$, the condition $\sigma_\omega\ll \mu P_0$ implies 
$\mu P_0 \gg |\omega'|$. This case then reduces to the previous one. 
In the corotating frame one has good adiabaticity and suppressed mixing, 
which means no wave packet separation and therefore no decoherence. Since 
decoherence is a physical process, it does not depend on the frame in which 
the evolution of the system is considered; therefore no decoherence occurs 
in the original flavour frame either. The individual flavour spin vectors 
$\vec{P}_\omega$ as well as the global flavour spin $\vec{P}$ 
are practically constant in the corotating frame, which means that 
in the original frame they all precess around $\vec{B}$ with the same 
frequency $\omega_s$, that is, synchronized oscillations occur. 

It is interesting to interpret this case also directly in the original 
flavour frame, without any reference to the corotating one. 
The condition $\sigma_\omega\ll \mu P_0 \lesssim \omega_0$ means that 
$\gamma \gtrsim 1$, that is adiabaticity is either moderately or 
strongly violated in the original frame.%
\footnote{Note that the degree of adiabaticity is 
frame-dependent because it is not a physically observable quantity. Only the 
probabilities of flavour transitions have direct physical meaning.} 
Strong violation of adiabaticity would mean that the propagation 
eigenstates are not a physically meaningful notion; even though they are 
mathematically well defined, they are strongly mixed and go into each 
other in the course of the evolution of the system. Group velocities are then 
not well defined either, and no wave packet separation 
can occur. To put it slightly differently, the initially 
produced propagation eigenstates with larger and smaller group velocities 
will fully (or almost fully) interchange on a time scale $\tau$ that is short 
compared to the naively expected coherence length $L_{\rm coh}$. The slow 
state becomes the fast one and vice versa; as a result, a small wave packet 
separation over the period $\tau$ is compensated during the next period $\tau$. 
As a consequence of frequent shuffling of the fast and slow propagation 
eigenstates, no noticeable wave packet separation occurs. 

\item[III.]
$\mu$ is slightly above $\mu_0\sim \sigma_\omega$. This case can be understood 
similarly to the previous one; the difference is that now 
 $\gamma \lesssim 1$, 
i.e.\ adiabaticity is only slightly violated. 
The shuffling of the fast and slow propagation eigenstate still occurs, but 
with an amplitude that is less than one: only a fraction of the fast 
propagation eigenstate goes into the slow one and vice versa.
As a result, at late enough times a split-up of the wave packets occurs, but 
the strengths of the separated wave packets 
are uneven: the probability of finding a neutrino in one of them is smaller 
than in the other. The wave packet separation is then only partial, which 
means that only partial decoherence has occurred. 
  
\item[IV.] $\mu < \mu_0$. In this case $\gamma = 0$ (perfect adiabaticity), 
so that the propagation eigenstates are well defined and evolve independently. 
The mixing angle at production is 
of order of vacuum mixing angle $\theta_0$, which means 
that the produced neutrino state is a nontrivial superposition of the two 
propagation eigenstates. At late times the complete wave packet separation 
occurs, leading to complete decoherence. 

\end{itemize}

More detailed discussion of the impact of adiabaticity and adiabaticity 
violation on decoherence of synchronized neutrino oscillations will be given 
in \cite{AKL4}.

\section{\label{sec:summary}Summary and discussion}
We have revisited synchronized neutrino oscillations in dense uniform and 
homogeneous neutrino gases. Our goal was twofold: (i) to give an approximate 
analytical description of the synchronized neutrino oscillations and of the 
de-synchronization phenomenon, and (ii) to interpret de-synchronization in 
terms of late-time decoherence. To this end, we first developed an exact 
formalism of spectral moments of the flavour spin vectors, and then 
applied it to find approximate analytical descriptions of decoherence effects 
in the system. Our spectral moments formalism also allowed us to find a 
previously unknown conservation law satisfied by the quantities characterizing 
a homogeneous and isotropic neutrino gas. 

In our first analytical approach, we assumed that at asymptotically 
large times the global flavour spin vector $\vec{P}$ undergoes a simple 
precession around the vector $\vec{B}$ that describes the vacuum 
contribution to the neutrino Hamiltonian, with the length of $\vec{P}$ 
remaining constant but possibly being smaller than its initial value 
$P_0$. The shrinkage of $\vec{P}$ signifies a partial or complete 
decoherence. We have found that this regime can only be nontrivially 
realized if the transverse (with respect to $\vec{B}$) components of 
$\vec{P}$ and of the vector $\vec{S}$ defined in eq.~(\ref{eq:EoM2}) 
satisfy $\vec{S}_\perp=\omega_s\vec{P}_\perp$, where $\omega_s$ is the 
precession frequency. We have additionally assumed that the 
longitudinal components of $\vec{P}$ and $\vec{S}$ satisfy a similar 
relation, i.e.\ $\vec{S}_\parallel=\omega_s\vec{P}_\parallel$. This 
allowed us to find (in Section~\ref{sec:const}) simple analytical 
expressions for the asymptotic value of $P_\perp$ and the precession 
frequency $\omega_s$ as functions of the neutrino self-interaction 
parameter $\mu$. In particular, we found a simple and direct 
mathematical explanation of the existence of the threshold value 
$\mu_0$, below which the complete decoherence occurs: for $\mu<\mu_0$ 
the quadratic equation from which $P=|\vec{P}|$ is determined has no 
real solutions. We thus confirmed the conclusions of 
Ref.~\cite{RaffTamb}, where the existence of the threshold $\mu_0$ and 
the threshold behaviour of the asymptotic value of $P_\perp$ were 
previously found.
  
The described approach, in addition to explaining the existence of the 
threshold $\mu_0$ and providing the correct order-of-magnitude estimate 
of its value ($\mu_0\sim \sigma_\omega$, where $\sigma_\omega$ is the 
effective width of the neutrino spectrum in the variable $\omega=\Delta 
m^2/(2p)$), predicted the correct behaviour of asymptotic $P_\perp$ at 
$\mu\gg \mu_0$. It, however, failed to reproduce the correct values of 
$P_\perp$ near the threshold $\mu_0$. One possible reason for this can 
be traced back to the fact that our additional assumption 
$\vec{S}_\parallel=\omega_s\vec{P}_\parallel$ is actually satisfied with 
a good accuracy far above the threshold but breaks down close to it. We 
therefore attempted to find an analytic approach based solely on the 
condition $\vec{S}_\perp=\omega_s\vec{P}_\perp$, which directly follows 
from the assumption of constant asymptotic $P$. By making use of our 
spectral moments formalism, we then found that this assumption cannot be 
exact as it leads to controversial results, at least for $\mu$ only 
slightly above the threshold $\mu_0$. This has been confirmed by our 
numerical calculations, which showed that for these values of $\mu$ both 
$P$ and $P_\perp$ do not become constant even at extremely late times, 
far above the naively expected coherence time $t_{\rm coh}\simeq L_{\rm 
coh}$. Instead, they continue oscillating around their mean values, 
though with relatively small amplitudes (see Fig.~\ref{fig:Plong}).

The smallness of the late-time oscillations of $P$ and $P_\perp$ suggests 
that they may be replaced at asymptotic times by their average values. 
However, as our analysis shows, this cannot be done by simply assuming 
them to be constant; instead, a consistent averaging procedure must be 
employed. We performed such a procedure in Section~\ref{sec:aver} 
by averaging the EoMs of the flavour spin vectors 
$\vec{P}_\omega$ over a very large time interval. The averaging is most simply 
done in the corotating frame, i.e.\ in the frame rotating around 
$\vec{B}$ with the angular velocity $\omega_s$ (see the Appendix). Our 
analysis in Section~\ref{sec:aver} was based on two simple observations:
\begin{itemize}
\item 
Large-interval time averages are dominated by the late-time behaviour 
of the system, so that by studying such averages one gains information 
about the asymptotic regime of evolution of the system. 

\item 
The smallness of the amplitude of the late-time oscillations of $P_\perp(t)$ 
means that in certain time integrals related to the averaging procedure 
$P_\perp(t)$ can be replaced by its asymptotic average value and pulled out 
of the integral.
\end{itemize}

\noindent
No further assumptions or approximations were used. 

The developed approach led to very simple analytical expressions 
for the averaged transverse and longitudinal components 
of the vector $\vec{P}_\omega$ in terms of the 
asymptotic value of $P_\perp$ and $\omega_s$, which can then be found 
from consistency conditions. Our numerical calculations demonstrate that 
the asymptotic $P_\perp$ and $\omega_s$ thus obtained reproduce very well the 
results of direct numerical integration of exact EoMs for $\vec{P}_\omega$ 
(see Fig.~\ref{fig:comparison}). 

The analytic expressions obtained in Section~\ref{sec:aver} coincide with 
those found in Ref.~\cite{RaffTamb} basing on different approximations and 
assumptions. Those assumptions are in general not realistic and, as was 
pointed out by the authors of \cite{RaffTamb} themselves, 
may actually be badly violated. 
How can then one understand the success of the analytical approach of 
\cite{RaffTamb} for describing the asymptotic values of $P_\perp$ and 
$\omega_s$?
As follows from our analysis, this is a consequence of cancellation of two 
large errors. The authors of \cite{RaffTamb} used the `sudden 
approximation' and the assumption that the values of the vectors 
$\vec{P}_\omega$ at the onset of the asymptotic regime coincide with their 
initial values at $t=0$ (see the discussion in our Section~\ref{sec:previous}). 
This allowed them to make the following replacements at asymptotic times 
(in our notation): 
\be
P_{\omega\parallel}(t)\to P_{\omega\parallel}(0)\,,\qquad
\vec{P}_\perp(t)\!\cdot\!\vec{P}_{\omega\perp}(t) \to P_\perp(t)
P_{\omega\perp}(0)\,,
\label{eq:repl1}
\ee
and as a result to replace  
\be
(\omega-\omega_r)P_{\omega\parallel}(t)+\mu 
\vec{P}_\perp(t)\!\cdot\!\vec{P}_{\omega\perp}(t) \to 
P_0 g_\omega\big[-c_{20}(\omega-\omega_r)+s_{20}\mu P_\perp(t)\big] 
\label{eq:repl2}
\ee
(cf.\ the numerators of the integrands in eqs.~(16) and~(17) of 
\cite{RaffTamb}). 
As follows from the calculations presented in the Appendix of our paper, 
although each of the two replacements in eq.~(\ref{eq:repl1}) is 
unjustified, the errors introduced by these replacements in the expression 
on the left hand side of eq.~(\ref{eq:repl2}) nearly cancel each other by 
virtue of the relation in eq.~(\ref{eq:correl1}), which is a direct consequence 
of the EoM for $\vec{P}_\omega$ in the corotating frame. 
Thus, our results provide a justification of the analytical 
approach of Ref.~\cite{RaffTamb}.

As was pointed out in Section~\ref{sec:previous}, the analytical expressions 
first obtained in \cite{RaffTamb} and rederived in a different approach 
in our Section~\ref{sec:aver} reproduce very well 
the results of numerical calculations of the asymptotic values of $P_\perp$, 
but they fail to satisfy the conservation laws (\ref{eq:Etot2}) and 
(\ref{eq:cons2a}) (except in the limit $\mu\gg\mu_0$). It is actually easy to 
understand why this happens. 
 The derivations of these expression involved replacing  the 
vectors $\vec{P}_\omega$ by their averages $\langle\vec{P}_\omega\rangle$. 
This allows an accurate determination of the averaged values of the 
longitudinal and transverse components of the global flavour spin vector 
$\vec{P}$, which are connected to the corresponding components of 
$\langle\vec{P}_\omega\rangle$ by linear relationships, see 
eq.~(\ref{eq:relations}). At the same time, the conservation laws 
(\ref{eq:Etot2}) and (\ref{eq:cons2a}) contain, along with linear members, 
terms that are quadratic or bilinear in the 
components of $\vec{P}$ and $\vec{S}$; their averages are not connected to the 
components of $\langle\vec{P}_\omega\rangle$ by linear relationships and 
therefore cannot be reliably found from the latter. 
The reason for this is essentially that the average of a square is not 
in general equal to the square of an average. On the other hand, in the limit 
$\mu\to \infty$ the condition $P_\perp=const.$ is very well satisfied 
at all times; in this case it is not necessary to invoke any averaging 
procedure, 
and the formulas derived for the averaged individual-mode and global flavour 
spin vectors 
are actually valid for the 
un-averaged quantities as well.  
The conservation laws (\ref{eq:Etot2}) and (\ref{eq:cons2a}) are then 
satisfied. 
Indeed, using eqs.~(\ref{eq:Pomegas}) and (\ref{eq:f1}) one can readily make 
sure that for general $\mu\ge \mu_0$ the differences of the left-hand and 
right-hand sides of eqs.~(\ref{eq:Etot2}) and (\ref{eq:cons2a}) 
 are proportional to $P_\perp-s_{20}P_0=P_\perp-P_\perp(0)$.
In the limit $\mu\to\infty$ this quantity vanishes, and the conservation 
laws (\ref{eq:Etot2}) and (\ref{eq:cons2a}) are fulfilled. 

In our analysis we had in mind a system of identical neutrino wave packets, 
with the function $g_\omega$ characterizing the energy distribution within 
each individual wave packet. Such a system is known to be equivalent to a 
system consisting of neutrinos with well-defined energy and $g_\omega$ 
characterizing the energy spectrum of the neutrino ensemble 
\cite{Kiers}. Our treatment therefore applies to such a system as well. 

Decoherence in a system of wave packets can be considered both in the 
momentum space and in the configuration space. In the 
latter case it is 
a consequence of separation of wave packets 
moving with different group velocities. In Section~\ref{sec:interp} we 
presented a qualitative interpretation of our results in terms of the 
possible wave packet separation. We have shown that all the regimes that we 
studied (perfect coherence and partial or full decoherence) can be understood 
from this standpoint. Our qualitative analysis there, however, did not provide 
a simple explanation of the existence of 
the threshold $\mu_0$, 
though it explained the complete decoherence for 
$\mu<\mu_0$ as being due to perfect adiabaticity. 

A dense uniform and isotropic neutrino gas that we considered in the 
present paper is the simplest possible system in which collective 
neutrino oscillations can occur. It can probably only very approximately 
represent the phenomena occurring in dense neutrino gases in the early 
Universe and to some extent in supernovae. Despite its simplicity, 
neutrino flavour evolution in this system exhibits a rich variety of 
possible patterns, the reason being that the equations of motion 
governing its evolution are highly nonlinear. We have addressed a number 
of issues pertaining to decoherence effects in this system. 
Some topics were, however, left out of our discussion. Those 
include e.g.\ the questions of what determines the relaxation time (i.e.\ 
the time necessary for the asymptotic regime to set in) and the amplitude 
of the residual oscillations of $P$ at late times. Hopefully, future studies 
will address these issues as well as will shed light on decoherence 
effects in collective neutrino oscillations in more realistic settings.

\section*{Acknowledgements} 
The authors are grateful to Baha Balantekin, 
Rasmus Lundkvist, Georg Raffelt, Alexei Smirnov and Cristina Volpe for very 
useful discussions and to Joachim Kopp for collaboration at an early stage of 
this work. Helpful correspondence with Georg Raffelt and Irene Tamborra is 
gratefully acknowledged.  
The work of A.M. is supported by the Italian Ministero 
dell'Istruzione, Universit\`a e Ricerca (MIUR) and Istituto Nazionale di 
Fisica Nucleare (INFN) through the ``Theoretical Astroparticle Physics'' 
projects.

\appendix
\renewcommand{\theequation}{\thesection\arabic{equation}}
\appsection
\renewcommand{\thesection}{\Alph{section}}
\section*{Appendix \Alph{section}:
Formalism in the corotating frame
and the \hspace*{3.8cm} evolution at asymptotically large times 
}

Consider the evolution of the flavour spin vectors 
in the corotating 
frame (i.e.\ in the frame rotating around $\vec{B}$ with the angular velocity 
$\omega_s$). We will mark the flavour spins and related quantities in this 
frame with a prime. 
Note that going to the 
corotating frame does not change the longitudinal (with respect to $\vec{B}$) 
components of the flavour spin vectors and spectral moments as well as the 
lengths of their transverse components. 
We shall be assuming that at $t=0$ the corotating frame coincides with the 
original one, so that the initial conditions for all the primed quantities are 
the same as for the corresponding unprimed ones. 

The EoM of the flavour spin vectors of the individual modes in the 
corotating frame is 
\be
\dot{\vec{P}}_\omega'=\vec{H}_\omega'\times\vec{P}_\omega'\,,
\label{eq:corEoM1}
\ee
where 
\be
\vec{H}_\omega'=(\omega-\omega_s)\vec{B}+\mu \vec{P}'=(\omega-\omega_r)\vec{B}
+\mu \vec{P}_\perp'\,.
\label{eq:corvecH1}
\ee 
Here the quantity $\omega_r$ was defined in eq.~(\ref{eq:omegar}), and 
in the last equality we have taken into account that the longitudinal 
component of $\vec{P}'$ is conserved and coincides with $\vec{P}_{\parallel}=
-c_{20}P_0\vec{B}$. 
The evolution equations for the longitudinal and transverse components of 
$\vec{P}_\omega'$ read 
\bea
&&\dot{P}_{\omega\parallel}'=\vec{B}\!\cdot\!(\mu\vec{P}_\perp'\times
\vec{P}_{\omega\perp}')\,,
\label{eq:corLong1} \\
&&\dot{\vec{P}}_{\omega\perp}'=\vec{B}
\times[(\omega-\omega_r)\vec{P}_{\omega\perp}'-\mu P_{\omega\parallel}'
\vec{P}_{\perp}']\,.
\label{eq:corTr1}
\eea
In eq.~(\ref{eq:corTr1}) the first term in the square brackets describes 
the precession of $\vec{P}_{\omega\perp}'$ around $\vec{B}$,
whereas the second term 
is responsible for the variations of the length of $\vec{P}_{\omega\perp}'$.  
Integrating eq.~(\ref{eq:corEoM1}) over $\omega$, we obtain the EoM of 
$\vec{P}'$: 
\be
\dot{\vec{P}}'=\vec{B}\times(\vec{S}_\perp'-\omega_s\vec{P}_\perp')\,.
\label{eq:corEoM2}
\ee
The EoMs of the spectral moments $\vec{K}_n'$ can be found by 
multiplying (\ref{eq:corEoM1}) by $\omega^n$ and integrating over $\omega$.

A useful relation is obtained by noting that eq.~(\ref{eq:corEoM1}) 
implies $\vec{H}_\omega'\!\cdot\!\dot{\vec{P}}_\omega'=0$, that is 
\be
(\omega-\omega_r)\dot{P}_{\omega\parallel}'+\mu\vec{P}_\perp'\!\cdot\!
\dot{\vec{P}}_{\omega\perp}'=0\,.
\label{eq:correl1}
\ee
Multiplying this by $\omega^n$ and integrating over $\omega$, one can find  
a scalar relation between the derivatives of $\vec{K}_n'$ and 
$\vec{K}_{n+1}'$, analogous to eq.~(\ref{eq:Kndiff2}). 

Consider now the asymptotic regime. 
If not indicated otherwise, in what follows we will be considering
all the relevant quantities at asymptotically large times, without
specifying this explicitly 
and keeping the same notation for these quantities as before.

Let us first assume that the evolution 
of the flavour spin vector $\vec{P}$ at late times is the simple precession 
around $\vec{B}$ with a constant frequency $\omega_s$. This means that in the 
corotating frame the flavour spin vector $\vec{P}'$ remains constant at 
asymptotic times. Eq.~(\ref{eq:corEoM2}) then yields
\be
\vec{S}_\perp'=\omega_s\vec{P}_\perp'\,,
\label{eq:corS3}
\ee
 which, in particular, means that $\vec{S}_\perp'$ is also 
constant at asymptotic times. At the same time, eq.~(\ref{eq:Etot1}) 
together with the asymptotic constancy of $P=P'$ implies $S_\parallel'=const$. 
Thus, we conclude that the vector $\vec{S}'$ is constant at late times. 

Acting by induction, it is then easy to show that all the spectral moments 
$\vec{K}_n'$ satisfy similar relations, 
that is, at asymptotically large times they all remain constant, with their 
transverse components collinear with $\vec{P}_\perp'$. (Note that such a 
behaviour in the corotating frame means that in the original flavour frame  
they all precess around $\vec{B}$ with the same constant frequency $\omega_s$, 
remaining in the same plane). 
{}From the definition of the spectral moments $\vec{K}_n'$, it then follows 
that the flavour spin vectors $\vec{P}_\omega'$ of the individual 
$\omega$-modes also satisfy similar relations, that is 
\be
\vec{P}_\omega'=const.,\qquad \vec{P}_{\omega\perp}'=a_\omega\vec{P}_\perp'
\,,
\label{eq:corPomega}
\ee
with constant $a_\omega$. By making use of eqs.~(\ref{eq:corPomega}) and 
(\ref{eq:corTr1}), one can find 
the ratio of the longitudinal and transverse components of the asymptotic 
vector $\vec{P}_\omega'$. 
The same relation will also hold for the corresponding unprimed quantities in 
the original flavour frame.  
Once the ratio of $P_{\omega\parallel}$ and $P_{\omega\perp}$ is known, one can 
then find these quantities from the normalization condition 
$P_{\omega\parallel}^2+P_{\omega\perp}^2=P_\omega^2=P_0^2g_\omega^2$.

With the expressions for $P_{\omega\parallel}$ and $P_{\omega\perp}$ at hand, 
the longitudinal and transverse components of $\vec{P}$ 
can be found by integrating over the $\omega$-modes. However, the direct 
calculation shows that, except in the limit $\mu\to\infty$, the 
 values of $P_\perp$ obtained in this way do not reproduce correctly the 
results of numerical integration of the exact EoMs. 
This means that the assumption that $P\equiv|\vec{P}|$ 
becomes constant at asymptotically large times, on which our consideration 
here was thus far based, is actually incorrect. 

Indeed, it was demonstrated in Section~\ref{sec:better} that for $\mu$ 
exceeding the critical value $\mu_0$ (but not far above it), the 
quantities $P_\perp$ and $P$ do not become constant even at very late 
times; instead, 
they keep oscillating around their mean values. 
The amplitude of the late-time oscillations of $P_\perp=P_\perp'$ is 
relatively small, and therefore it seems to be a reasonable approximation 
to replace it at asymptotic times by its mean value, found by averaging over 
these oscillations. 
However, this should be done through a consistent averaging procedure 
rather than by simply assuming that at large times $P_\perp'=const$.

To carry out the averaging procedure systematically, 
let us average the EoMs of the flavour spin vectors over a very large 
(formally infinite) time interval. Since the infinite-interval average of the 
derivative of any bounded function vanishes,%
\footnote{
\label{fooot:avderiv}
Indeed, for a bounded function $f(t)$ one has  $\langle \dot{f}(t)\rangle\equiv
\lim\limits_{T\to\infty} \frac{1}{T}\int_0^T \dot{f}(t)dt=
\lim\limits_{T\to\infty} \frac{1}{T}[f(T)-f(0)]=0$.}	
eqs.~(\ref{eq:corLong1})-(\ref{eq:corEoM2}) yield 
\be
\langle\vec{P}_\perp'\times
\vec{P}_{\omega\perp}'\rangle=0\,,\hspace*{1.9cm}
\label{eq:corLong2} 
\ee
\be
(\omega-\omega_r)\langle\vec{P}_{\omega\perp}'\rangle=\mu 
\langle P_{\omega\parallel}'\vec{P}_{\perp}'\rangle\,,
\vspace*{1.5mm}
\label{eq:corTr2}
\ee
\be
\langle
\vec{S}_\perp'\rangle=\omega_s\langle\vec{P}_\perp'\rangle\,.\hspace*{2.15cm}
\label{eq:corEoM3}
\ee
Note that eq.~(\ref{eq:corEoM3}) is simply the averaged version of 
eq.~(\ref{eq:corS3}). 

To draw 
useful information
from eq.~(\ref{eq:corTr2}), let us note that, while  
at late times the transverse component of the {\sl global} flavour spin 
in the corotating frame $\vec{P}_\perp'$ is essentially a fixed-direction 
vector with its length exhibiting only small oscillations, this is in general 
not the case for the components of the flavour spin vectors $\vec{P}_\omega'$ 
of the {\sl individual} $\omega$-modes, which undergo large-scale variations.%
\footnote{
\label{foot:fullscale}
This can be seen from eq.~(\ref{eq:corTr1}). For instance, in the 
limit $\mu\to\mu_0$ we have $P_\perp'\to 0$, and the second term in the square 
brackets in this equation vanishes, whereas the first term describes the 
undamped precession of $\vec{P}_\perp'$ around $\vec{B}$ in the plane 
perpendicular to $\vec{B}$ with the frequency $\omega-\omega_r$. For 
$\mu>\mu_0$ also the length of $\vec{P}_{\omega\perp}'$ varies, and therefore 
so does $P_{\omega\parallel}'$.
}  
Because infinite-time averages are dominated by the late-time contributions 
to the averaging integral, it is then a good approximation to 
replace the nearly constant vector $\vec{P}_\perp'(t)$ 
by its mean value and pull it out of the integral 
when calculating the average of $P_{\omega\parallel}'\vec{P}_{\perp}'$. 
This gives%
\be
\langle P_{\omega\parallel}'\vec{P}_{\perp}'\rangle \simeq
\langle P_{\omega\parallel}'\rangle\langle\vec{P}_{\perp}'\rangle\,.
\label{eq:corTr2a}
\ee
We will be using the same factorization approximation whenever calculating 
the averages of the products of $\vec{P}_\perp'$ and any components of 
$\vec{P}_{\omega}'$. 

Substituting eq.~(\ref{eq:corTr2a}) into (\ref{eq:corTr2}) yields 
\be
\frac{\langle\vec{P}_{\omega\perp}'\rangle}
{\langle P_{\omega\parallel}'\rangle}=\frac{\mu\langle\vec{P}_{\perp}'\rangle}
{(\omega-\omega_r)}\,.
\label{eq:corTr2b}
\ee
Note that 
in the factorization approximation 
eq.~(\ref{eq:corLong2}) becomes $\langle\vec{P}_\perp'\rangle\times
\langle\vec{P}_{\omega\perp}'\rangle=0$. This relation does not bring in any 
new information, as it follows also from (\ref{eq:corTr2b}).

Eq.~(\ref{eq:corTr2b}) equips us with the ratio of 
$\langle\vec{P}_{\omega\perp}'\rangle$ and $\langle P_{\omega\parallel}'
\rangle$. However, one cannot combine it with the normalization condition for 
$P_\omega'^{2}$ in order to find 
$\langle\vec{P}_{\omega\perp}'\rangle$ and $\langle P_{\omega\parallel}'
\rangle$ separately, since these averages do not satisfy the same 
normalization condition as 
the un-averaged quantities $\vec{P}_{\omega\perp}'$ and $P_{\omega\parallel}'$. 
Therefore, in order to determine 
$\langle P_{\omega\parallel}'\rangle$ and $\langle\vec{P}_{\omega\perp}'
\rangle$, one needs one more relation between them. 
To find it, we first rewrite eq.~(\ref{eq:corTr2b}) as 
\be
\langle \vec{P}_{\omega\perp}'\rangle=f(\omega,\mu)
\langle\mu\vec{P}_\perp\rangle\,,\qquad~
\nonumber 
\ee
\be
\langle P_{\omega\parallel}'\rangle=f(\omega,\mu)
\big[\omega-\omega_r(\mu)\big]\,,
\label{eq:corPomegas}
\ee
with as yet unknown function $f(\omega,\mu)$.
Next, consider the time average of the quantity $\vec{P}_\omega'\!\cdot\!
\vec{H}_\omega'$. Using eq.~(\ref{eq:corPomegas}) and the definition of 
$\vec{H}_\omega'$ given in eq.~(\ref{eq:corvecH1}), one can express this 
average through $f(\omega,\mu)$: 
\be
\langle\vec{P}_\omega'\!\cdot\!\vec{H}_\omega'\rangle=
f(\omega,\mu)\big[(\omega-\omega_r)^2+\langle\mu P_\perp\rangle^2\big].
\label{eq:PH1}
\ee
In obtaining this relation we have 
used the factorization approximation for 
$\langle\vec{P}_{\omega\perp}'\!\cdot\!\vec{P}_{\perp}'
\rangle$. 
On the other hand, at a given time $t_1$, for the un-averaged quantity  
$\vec{P}_\omega'(t_1)\!\cdot\!\vec{H}_\omega'(t_1)$ one can write 
\be
\vec{P}_\omega'(t_1)\!\cdot\!\vec{H}_\omega'(t_1)=
(\omega-\omega_r)P_{\omega\parallel}'(t_1)+
\mu\vec{P}_{\perp}'(t_1)\!\cdot\!\vec{P}_{\omega\perp}'(t_1) 
\hspace*{6.5cm}
\vspace*{-1.5mm}
\nonumber 
\ee
\be
=(\omega-\omega_r)P_{\omega\parallel}'(0)+
\mu\vec{P}_\perp'(t_1)\!\cdot\!\vec{P}_{\omega\perp}'(0)
\hspace*{4.4cm}
\vspace*{0.5mm}
\nonumber 
\ee
\be
\hspace*{1.7cm}+
\big\{(\omega-\omega_r)[P_{\omega\parallel}'(t_1)-P_{\omega\parallel}'(0)]
+\mu\vec{P}_{\perp}'(t_1)\!\cdot\!\big[\vec{P}_{\omega\perp}'(t_1)
-\vec{P}_{\omega\perp}'(0)\big]
\big\}\,.
\label{eq:longcalc1}
\ee
Let us now consider the expression in the curly brackets in 
(\ref{eq:longcalc1}). 
\noindent
We have 
\be
\big\{(\omega-\omega_r)[P_{\omega\parallel}'(t_1)-P_{\omega\parallel}'(0)]
+\mu\vec{P}_{\perp}'(t_1)\!\cdot\!\big[\vec{P}_{\omega\perp}'(t_1)
-\vec{P}_{\omega\perp}'(0)\big]
\big\}
\hspace*{1.2cm}
\nonumber
\ee
\be
=\int_0^{t_1} dt\,\big[(\omega-\omega_r)\dot{P}_{\omega\parallel}'(t)
+\mu\vec{P}_{\perp}'(t)\!\cdot\!\dot{\vec{P}}_{\omega\perp}'(t)
-\mu\vec{P}_{\perp}'(t)\!\cdot\!\dot{\vec{P}}_{\omega\perp}'(t)\big]
\nonumber 
\ee
\be
+\mu\vec{P}_{\perp}'(t_1)\!\cdot\!\big[\vec{P}_{\omega\perp}'(t_1)
-\vec{P}_{\omega\perp}'(0)\big]
\nonumber 
\ee
\be
\hspace*{-0.45cm}
=-\mu\int_0^{t_1} dt\,\vec{P}_{\perp}'(t)\!\cdot\!\dot{\vec{P}}_{\omega\perp}'
(t)
+\mu\vec{P}_{\perp}'(t_1)\!\cdot\!\big[\vec{P}_{\omega\perp}'(t_1)
-\vec{P}_{\omega\perp}'(0)\big]\,.
\label{eq:longcalc2}
\ee
Here in going from lines 2, 3 to line 4 we have used 
relation~(\ref{eq:correl1}).
We will eventually be interested in the time average of 
eq.~(\ref{eq:longcalc2}). Since the integrals related to infinite-interval 
averages are dominated by the contributions of late times in the integration 
intervals, we can concentrate on the large $t_1$ limit of (\ref{eq:longcalc2}). 
The integral in the last line of this equation is also dominated by the 
late-time contributions,
as its integrand is not suppressed at large $t$. 
Since $\vec{P}_{\perp}'(t)$ is nearly constant at asymptotic times, one can 
approximately replace it by its value at $t=t_1$ and pull it out of the 
integral. The two terms in last line in eq.~(\ref{eq:longcalc2}) 
then cancel each other, which means that  
the average of the 
expression in the curly brackets in 
(\ref{eq:longcalc1}) can be neglected. Thus, the averaging of 
eq.~(\ref{eq:longcalc1}) yields
\be
\langle\vec{P}_\omega'\!\cdot\!\vec{H}_\omega'\rangle\simeq
(\omega-\omega_r)P_{\omega\parallel}'(0)+\vec{P}_{\omega\perp}'(0)
\!\cdot\!\langle\mu\vec{P}_{\perp}'\rangle\,. 
\label{eq:HPav1}
\ee
The quantities $\vec{P}_{\omega\perp}'(0)$ are constant vectors in the 
corotating frame which lie in the $x'z'$ plane and all point in the same 
direction, irrespectively of the value of $\omega$. 
This follows from the fact that at $t=0$ they coincide with 
the corresponding $\vec{P}_{\omega\perp}$, and the initial conditions for 
$\vec{P}_\omega$ in eq.~(\ref{eq:init1}) mean that all 
$\vec{P}_{\omega\perp}(0)$ point in the same direction. 
The averaged global flavour spin $\langle \vec{P}_{\perp}'\rangle$ is also a 
constant vector in the corotating frame. Its direction can only depend 
on the vectors characterizing the neutrino system under consideration and 
should be given by their linear superposition. Those are the vector 
$\vec{n}_z'(0)=\vec{n}_z$, which specifies the initial conditions for the 
flavour spin vectors, and $\vec{B}$. The latter, as well as the longitudinal 
component of the former, cannot enter in the definition of a transverse vector, 
whereas the transverse component of $\vec{n}_z'(0)$ 
defines the direction of $\vec{P}_{\omega\perp}'(0)$. We therefore conclude 
that $\langle \vec{P}_{\perp}'\rangle$ and $\vec{P}_{\omega\perp}'(0)$ 
must be  collinear.%
Thus, one can rewrite eq.~(\ref{eq:HPav1}) as 
\be
\langle\vec{P}_\omega'\!\cdot\!\vec{H}_\omega'\rangle\simeq
(\omega-\omega_r)P_{\omega\parallel}'(0)+\mu\langle P_{\perp}'\rangle
P_{\omega\perp}'(0)\,\;\nonumber \qquad\qquad~
\vspace*{-1.2mm}
\ee
\be
=\big[-c_{20}(\omega-\omega_r)+s_{20}\mu \langle P_{\perp}
\rangle\big] P_0 g_\omega\,.
\label{eq:HPav}
\ee
Here we have taken into account that the initial conditions for the primed 
and the corresponding unprimed components of the flavour spin vectors 
are the same and that $P_\perp'=P_\perp$. 
Combining eqs.~(\ref{eq:HPav}) and (\ref{eq:PH1}), one arrives at 
the expression for $f(\omega,\mu)$ given in eq.~(\ref{eq:f1}).

\end{document}